# Error Characterization, Mitigation, and Recovery in Flash-Memory-Based Solid-State Drives

Yu Cai†     Saugata Ghose†     Erich F. Haratsch‡     Yixin Luo†     Onur Mutlu§†

†Carnegie Mellon University     ‡Seagate Technology     §ETH Zürich

## ABSTRACT

NAND flash memory is ubiquitous in everyday life today because its capacity has continuously increased and cost has continuously decreased over decades. This positive growth is a result of two key trends: (1) effective process technology scaling; and (2) multi-level (e.g., MLC, TLC) cell data coding. Unfortunately, the reliability of raw data stored in flash memory has also continued to become more difficult to ensure, because these two trends lead to (1) fewer electrons in the flash memory cell floating gate to represent the data; and (2) larger cell-to-cell interference and disturbance effects. Without mitigation, worsening reliability can reduce the lifetime of NAND flash memory. As a result, flash memory controllers in solid-state drives (SSDs) have become much more sophisticated: they incorporate many effective techniques to ensure the correct interpretation of noisy data stored in flash memory cells. In this article, we review recent advances in SSD error characterization, mitigation, and data recovery techniques for reliability and lifetime improvement. We provide rigorous experimental data from state-of-the-art MLC and TLC NAND flash devices on various types of flash memory errors, to motivate the need for such techniques. Based on the understanding developed by the experimental characterization, we describe several mitigation and recovery techniques, including (1) cell-to-cell interference mitigation; (2) optimal multi-level cell sensing; (3) error correction using state-of-the-art algorithms and methods; and (4) data recovery when error correction fails. We quantify the reliability improvement provided by each of these techniques. Looking forward, we briefly discuss how flash memory and these techniques could evolve into the future.

## 1 INTRODUCTION

Solid-state drives (SSDs) are widely used in computer systems today as a primary method of data storage. In comparison with magnetic hard drives, the previously dominant choice for storage, SSDs deliver significantly higher read and write performance, with orders of magnitude of improvement in random-access input/output (I/O) operations, and are resilient to physical shock, while requiring a smaller form factor and consuming less static power. SSD capacity (i.e., storage density) and cost-per-bit have been improving steadily in the past two decades, which has led to the widespread adoption of SSD-based data storage in most computing systems, from mobile consumer devices [65, 76] to enterprise data centers [51, 122, 142, 162, 178].

The first major driver for the improved SSD capacity and cost-per-bit has been *manufacturing process scaling*, which has increased the number of flash memory cells within a fixed area. Internally, commercial SSDs are made up of NAND flash memory chips, which provide nonvolatile memory storage (i.e., the data stored in NAND flash is correctly retained even when the power is disconnected) using *floating gate (FG) transistors* [77, 121, 130] or *charge trap transistors* [50, 186]. In this paper, we mainly focus on floating gate transistors, since they are the most common transistor used in today's flash memories. A floating gate transistor constitutes a flash memory cell. It can encode one or more bits of digital data, which is represented by the level of charge stored inside the transistor's *floating gate*. The transistor traps charge within its floating gate, which dictates the *threshold voltage* level at which the transistor turns on. The threshold voltage level of the floating gate is used to determine the value of the digital data stored inside the transistor. When manufacturing process scales down to a smaller technology node, the size of each flash memory cell, and thus the size of the transistor, decreases, which in turn reduces the amount of charge that can be trapped within the floating gate. Thus, process scaling increases storage density by enabling more cells to be placed in a given area, but it also causes reliability issues, which are the focus of this paper.

The second major driver for improved SSD capacity has been the use of a single floating gate transistor to represent *more than* one bit of digital data. Earlier NAND flash chips stored a single bit of data in each cell (i.e., a single floating gate transistor), which was referred to as single-level cell (SLC) NAND flash. Each transistor can be set to a specific threshold voltage within a fixed range of voltages. SLC NAND flash divided this fixed range into two *voltage windows*, where one window represents the bit value 0 and the other window represents the bit value 1. Multi-level cell (MLC) NAND flash was commercialized in the last two decades, where the same voltage range is instead divided into *four* voltage windows that represent each possible 2-bit value (00, 01, 10, and 11). Each voltage window in MLC NAND flash is therefore much smaller than a voltage window in SLC NAND flash. This makes it more difficult to identify the value stored in a cell. More recently, triple-level cell (TLC) flash has been commercialized [4, 62], which further divides the range, providing *eight* voltage windows to represent a 3-bit value. Quadruple-level cell (QLC) flash, storing a 4-bit value per cell, is currently being developed [145]. Encoding more bits per cell increases the capacity of the SSD without increasing the chip size, yet it also decreases reliability by making it more difficult to correctly store and read the bits.

The two major drivers for the higher capacity, and thus the ubiquitous commercial success, of flash memory as a storage device, are also major drivers for its reduced reliability and are the causes of its scaling problems. As the amount of charge stored in each NAND flash cell decreases, the voltage for each possible bit value is distributed over a wider voltage range due to greater process variation, and the *margins* (i.e., the width of the gap between neighboring voltage windows) provided to ensure the raw reliability of NAND flash chips have been diminishing, leading to a greater probability of flash memory errors with newer generations of SSDs. NAND flash memory errors can be induced by a variety of sources [14], including flash cell wearout [14, 15, 116], errors introduced during programming [12, 18, 116, 153], interference from operations performed on adjacent cells [16, 18, 26, 56, 108, 126, 149, 151], and data retention issues due to charge leakage [14, 17, 24, 25, 126].

To compensate for this, SSDs employ sophisticated error-correcting codes (ECCs) within their controllers. An SSD controller uses the ECC information stored alongside a piece of data in the NAND flash chip to detect and correct a number of *raw bit errors* (i.e., the number of errors experienced before correction is applied) when the piece of data is read out. The number of bits that can be corrected for every piece of data is a fundamental tradeoff in an SSD. A more sophisticated ECC can tolerate a larger number of raw bit errors, but it also consumes greater area overhead and latency. Error characterization studies [14, 15, 56, 116, 126, 153] have found that, due to NAND flash wearout, the probability of raw bit errors increases as more *program/erase (P/E) cycles* (i.e., *write accesses*, or *writes*) are performed to the drive. The raw bit error rate eventually exceeds the maximum number of errors that can be corrected by ECC, at which point data loss occurs [17, 22, 122, 162]. The *lifetime* of a NAND-flash-memory-based SSD is determined by the number of P/E cycles that can be performed successfully while avoiding data loss for a minimum *retention guarantee* (i.e., the required minimum amount of time, after being written, that the data can still be read out without uncorrectable errors).

The decreasing raw reliability of NAND flash memory chips has drastically impacted the lifetime of commercial SSDs. For example, older SLC NAND-flash-based SSDs were able to withstand 150,000 P/E cycles (writes) to each flash cell, but contemporary 1x-nm (i.e., 15–19 nm) process-based SSDs consisting of MLC NAND flash can sustain only 3,000 P/E cycles [120, 153, 206]. With the raw reliability of a flash chip dropping so significantly, approaches to mitigating reliability issues in NAND-flash-based SSDs have been the focus of an important body of research. A number of solutions have been proposed to increase the lifetime of contemporary SSDs, ranging from changes to the low-level device behavior (e.g., [12, 15, 16, 201]) to making SSD controllers much more intelligent in dealing with individual flash memory chips (e.g., [17, 21, 23–26, 62, 115, 116]). In addition, various mechanisms have been developed to successfully recover data in the event of data loss that may occur during a read operation to the SSD (e.g., [16, 17, 21]).

In this work, we provide a comprehensive overview of the state of flash-memory-based SSD reliability, with a focus on (1) fundamental causes of flash memory errors, backed up by (2) quantitative error data collected from real state-of-the-art flash memory devices, and (3) sophisticated error mitigation and data recovery techniques developed to tolerate, correct, and recover from such errors. To this end, we first discuss the architecture of a state-of-the-art SSD, and describe mechanisms used in a commercial SSD to reduce the probability of data loss (Section 2). Next, we discuss the low-level behavior of the underlying NAND flash memory chip in an SSD, to illustrate fundamental reasons why errors can occur in flash memory (Section 3). We then discuss the root causes of these errors, quantifying the impact of each error source using experimental characterization data collected from real NAND flash memory chips (Section 4). For each of these error sources, we describe various state-of-the-art mechanisms that mitigate the induced errors (Section 5). We next examine several error recovery flows to successfully extract data from the SSD in the event of data loss during a read operation (Section 6). Then, we look to the future to foreshadow how the reliability of SSDs might be affected by emerging flash memory technologies (Section 7). Finally, we briefly examine how other memory technologies (such as DRAM, which is used prominently in a modern SSD, and emerging nonvolatile memory) suffer from similar reliability issues to SSDs (Section 8).

## 2 STATE-OF-THE-ART SSD ARCHITECTURE

In order to understand the root causes of reliability issues within SSDs, we first provide an overview of the system architecture of a state-of-the-art SSD. The SSD consists of a group of NAND flash memories (or *chips*) and a *controller*, as shown in Figure 1. A host computer communicates with the SSD through a high-speed host interface (e.g., SAS, SATA, PCIe bus), which connects to the SSD controller. The controller is then connected to each of the NAND flash chips via memory *channels*.

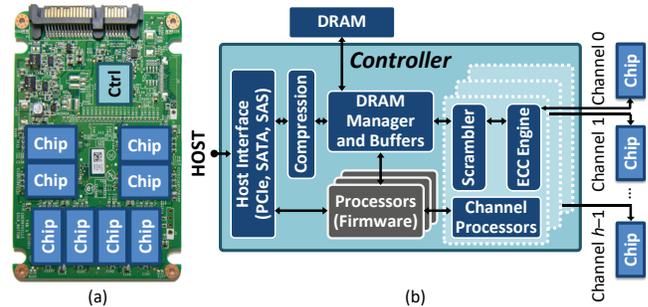

**Figure 1:** (a) SSD system architecture, showing controller (Ctrl) and chips. (b) Detailed view of connections between controller components and chips.

### 2.1 Flash Memory Organization

Figure 2 shows an example of how NAND flash memory is organized within an SSD. The flash memory is spread across multiple flash chips, where each chip contains one or more flash *dies*, which are individual pieces of silicon wafer that are connected together to the pins of the chip. Contemporary SSDs typically have 4–16 chips per SSD, and can have as many as 16 dies per chip. Each chip is connected to one or more physical memory channels, and these memory channels are not shared across chips. A flash die operates independently of other flash dies, and contains between one and four planes. Each plane contains hundreds to thousands of flash blocks. Each block is a 2D array that contains hundreds of rows of flash cells (typically 256–1024 rows) where the rows store contiguous pieces of data. Much like banks in a multi-bank memory (e.g., DRAM banks [31, 91, 92, 100, 102, 104, 105, 131, 137, 138]), the planes can execute flash operations in parallel, but the planes within a die share a single set of data and control buses [1]. Hence, an operation can be started in a different plane in the same die in a pipelined manner, every cycle. Figure 2 shows how blocks are organized within chips across multiple channels. In the rest of this work, without loss of generality, we assume that a chip contains a single die.

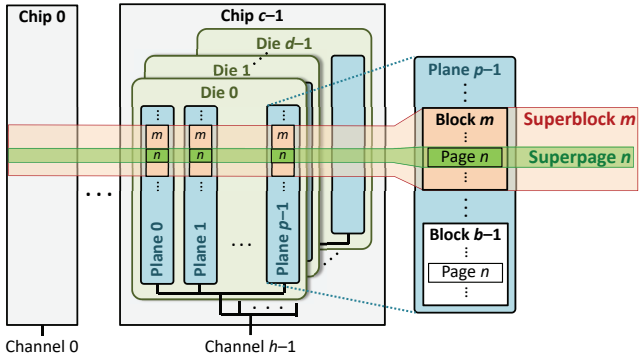

Figure 2: Flash memory organization.

Data in a block is written at the unit of a *page*, which is typically between 8 and 16 kB in size in NAND flash memory. All read and write operations are performed at the granularity of a page. Each block typically contains hundreds of pages. Blocks in each plane are numbered with an ID that is unique within the plane, but is shared across multiple planes. Within the block, each page is numbered in sequence. The controller firmware groups blocks with the same ID number across multiple chips and planes together into a *superblock*. Within each superblock, the pages with the same page number are considered a *superpage*. The controller *opens* one superblock (i.e., an empty superblock is selected for write operations) at a time, and typically writes data to the NAND flash memory one superpage at a time to improve sequential read/write performance and make error correction efficient, since some parity information is kept at superpage granularity (see Section 2.3). Having the ability to write to all of the pages in a superpage simultaneously, the SSD can fully exploit the internal parallelism offered by multiple planes/chips, which in turn maximizes write throughput.

## 2.2 Memory Channel

Each flash memory channel has its own data and control connection to the SSD controller, much like a main memory channel has to the DRAM controller [90, 91, 93, 132, 134, 137, 138, 173, 174]. The connection for each channel is typically an 8- or 16-bit wide bus between the controller and one of the flash memory chips [1]. Both data and flash commands can be sent over the bus.

Each channel also contains its own control signal pins to indicate the type of data or command that is on the bus. The *address latch enable* (ALE) pin signals that the controller is sending an address, while the *command latch enable* (CLE) pin signals that the controller is sending a flash command. Every rising edge of the *write enable* (WE) signal indicates that the flash memory should write the piece of data currently being sent on the bus by the SSD controller. Similarly, every rising edge of the *read enable* (RE) signal indicates that the flash memory should send the next piece of data from the flash memory to the SSD controller.

Each flash memory die connected to a memory channel has its own *chip enable* (CE) signal, which selects the die that the controller currently wants to communicate with. On a channel, the bus broadcasts address, data, and flash commands to all dies within the channel, but only the die whose CE signal is active reads the information from the bus and executes the corresponding operation.

## 2.3 SSD Controller

The SSD controller, shown in Figure 1b, is responsible for managing the underlying NAND flash memory, and for handling I/O requests received from the host. To perform these tasks, the controller runs firmware, which is often referred to as the *flash translation layer* (FTL). FTL tasks are executed on one or more embedded processors that exist inside the controller. The controller has access to DRAM, which can be used to store various controller metadata (e.g., how host memory addresses map to physical SSD addresses) and to cache relevant (e.g., frequently accessed) SSD pages [122, 161]. When the controller handles I/O requests, it performs a number of operations on the data, such as *scrambling* the data to improve raw bit error rates, performing *ECC encoding/decoding*, and in some cases *compressing* the data and employing *superpage-level data parity*. We briefly examine the various tasks of the SSD controller.

**Flash Translation Layer.** The main duty of the FTL is to manage the mapping of *logical addresses* (i.e., the address space utilized by the host) to *physical addresses* in the underlying flash memory (i.e., the address space for actual locations where the data is stored, visible only to the SSD controller) for each page of data [40, 58]. By providing this indirection between address spaces, the FTL can *remap* the logical address to a different physical address (i.e., move the data to a different physical address) *without* notifying the host. Whenever a page of data is written to by the host or moved for underlying SSD maintenance operations (e.g., garbage collection [35, 202]; see below), the old data (i.e., the physical location where the overwritten data resides) is simply marked as invalid in the physical block's *metadata*, and the new data is written to a page in the flash block that is currently open for writes (see Section 3.4 for more detail on how writes are performed).

Over time, page invalidations cause *fragmentation* within a block, where a majority of pages in the block become invalid. The FTL periodically performs *garbage* collection, which identifies each of the highly fragmented flash blocks and erases the entire block (after migrating any remaining valid pages to a new block, with the goal of fully populating the new block with valid pages) [35, 202]. Garbage collection often aims to select the blocks with the least amount of utilization (i.e., the fewest valid pages) first. When garbage collection is complete, and a block has been erased, it is added to a *free list* in the FTL. When the block currently open for writes becomes full, the SSD controller selects a new block to open from the free list.

The FTL is also responsible for *wear leveling*, to ensure that all of the blocks within the SSD are evenly worn out [35, 202]. By evenly distributing the *wear* (i.e., the number of P/E cycles that take place) *across* different blocks, the SSD controller reduces the heterogeneity of the amount of wearout across these blocks, extending the lifetime of the device. Wear-leveling algorithms are invoked when the current block that is being written to is full (i.e., no more pages in the block are available to write to), and the controller selects a new block for writes from the free list. The wear-leveling algorithm dictates which of the blocks from the free list is selected. One simple approach is to select the block in the free list with the lowest number of P/E cycles to minimize the variance of the wearout amount across blocks, though many algorithms have been developed for wear leveling [34, 54].

**Flash Reliability Management.** The SSD controller performs many background optimizations that improve flash reliability. These flash reliability management techniques, as we will discuss in more detail in Section 5, can effectively improve flash lifetime at a very low cost, since the optimizations are usually performed during idle times, when the interference with the running workload is minimized. These management techniques sometimes require small metadata storage in memory (e.g., for storing optimal read reference voltages [16, 17, 116]), or require a timer (e.g., for triggering refreshes in time [24, 25]).

**Compression.** Compression can reduce the size of the data written to minimize the number of flash cells worn out by the original data. Some controllers provide compression, as well as decompression, which reconstructs the original data from the compressed data stored in the flash memory [110, 209]. The controller may contain a *compression engine*, which, for example, performs the LZ77 or LZ78 algorithms. Compression is optional, as some types of data being stored by the host (e.g., JPEG images, videos, encrypted files, files that are already compressed) may not be compressible.

**Data Scrambling and Encryption.** The occurrence of errors in flash memory is highly dependent on the data values stored into the memory cells [14, 18, 26]. To reduce the dependence of the error rate on data values, an SSD controller first scrambles the data before writing it into the flash chips [27, 84]. The key idea of scrambling is to probabilistically ensure that the actual value written to the SSD contains an equal number of randomly distributed zeroes and ones, thereby minimizing any data-dependent behavior. Scrambling is performed using a reversible process, and the controller *descrambles* the data stored in the SSD during a read request. The controller employs a *linear feedback shift register* (LFSR) to perform scrambling and descrambling. An $n$-bit LFSR generates $2^{n-1}$ bits worth of pseudo-random numbers without repetition. For each page of data to be written, the LFSR can be seeded with the *logical* address of that page, so that the page can be correctly descrambled even if maintenance operations (e.g., garbage collection) migrate the page to another physical location, as the logical address is unchanged. (This also reduces the latency of maintenance operations, as they do not need to descramble and rescramble the data when a page is migrated.) The LFSR then generates a pseudo-random number based on the seed, which is then XORed with the data to produce the scrambled version of the data. As the XOR operation is reversible, the same process can be used to descramble the data.

In addition to the data scrambling employed to minimize data value dependence, several SSD controllers include data encryption hardware [41, 64, 189]. An SSD that contains data encryption hardware within its controller is known as a *self-encrypting drive* (SED). In the controller, data encryption hardware typically employs AES encryption [41, 45, 144, 189], which performs multiple rounds of substitutions and permutations to the unencrypted data in order to encrypt it. AES employs a separate key for each round [45, 144]. In an SED, the controller contains hardware that generates the AES keys for each round, and performs the substitutions and permutations to encrypt or decrypt the data using dedicated hardware [41, 64, 189].

**Error-Correcting Codes.** ECC is used to detect and correct the raw bit errors that occur within flash memory. A host writes a page of data, which the SSD controller splits into one or more chunks. For each chunk, the controller generates a *codeword*, consisting of the chunk and a correction code. The strength of protection offered by ECC is determined by the *coding rate*, which is the chunk size divided by the codeword size. A higher coding rate provides weaker protection, but consumes less storage, representing a key reliability tradeoff in SSDs.

The ECC algorithm employed (typically BCH [6, 66, 109, 168] or LDPC [55, 119, 168, 207]; see Section 6), as well as the length of the codeword and the coding rate, determine the total *error correction capability*, i.e., the maximum number of raw bit errors that can be corrected by ECC. ECC engines in contemporary SSDs are able to correct data with a relatively high raw bit error rate (e.g., between $10^{-3}$ and $10^{-2}$ [72]) and return data to the host at an error rate that meets traditional data storage reliability requirements (e.g., a post-correction error rate of $10^{-15}$ in the JEDEC standard [74]). The *error correction failure rate* ($P_{ECFR}$) of an ECC implementation, with a codeword length of $l$ where the codeword has an error correction capability of $t$ bits, can be modeled as:

$$P_{ECFR} = \sum_{k=t+1}^{l} \binom{l}{k}(1-\text{BER})^{(l-k)}\text{BER}^k \quad (1)$$

where BER is the bit error rate of the NAND flash memory. We assume in this equation that errors are independent and identically distributed.

In addition to the ECC information, a codeword contains cyclic redundancy checksum (CRC) parity information [161]. When data is being read from the NAND flash memory, there may be times when the ECC algorithm incorrectly indicates that it has successfully corrected all errors in the data, when uncorrected errors remain. To ensure that incorrect data is not returned to the user, the controller performs a CRC check in hardware to verify that the data is error free [157, 161].

**Data Path Protection.** In addition to protecting the data from raw bit errors within the NAND flash memory, newer SSDs incorporate error detection and correction mechanisms throughout the SSD controller, in order to further improve reliability and data integrity [161]. These mechanisms are collectively known as *data path protection*, and protect against errors that can be introduced by the various SRAM and DRAM structures that exist within the SSD.[1] Figure 3 illustrates the various structures within the controller that employ data path protection mechanisms. There are three data paths that require protection: (1) the path for data written by the host to the flash memory, shown as a red solid line in Figure 3; (2) the path for data read from the flash memory by the host, shown as a green dotted line; and (3) the path for metadata transferred

---
[1]See Section 8 for a discussion on the possible types of errors that can be present in DRAM.

between the firmware (i.e., FTL) processors and the DRAM, shown as a blue dashed line.

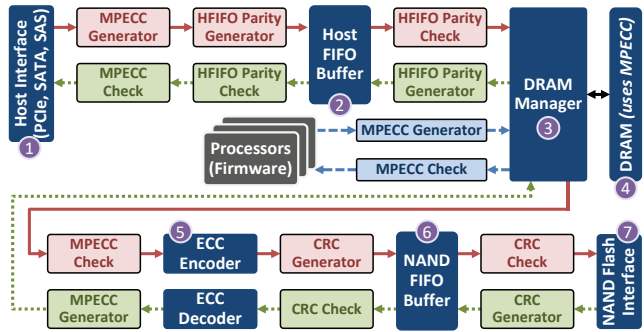

**Figure 3: Data path protection employed within the controller.**

In the write data path of the controller (the red solid line shown in Figure 3), data received from the host interface (❶ in the figure) is first sent to a host FIFO buffer (❷). Before the data is written into the host FIFO buffer, the data is appended with *memory protection ECC* (MPECC) and *host FIFO buffer* (HFIFO) parity [161]. The MPECC parity is designed to protect against errors that are introduced when the data is stored within DRAM (which takes place later along the data path), while the HFIFO parity is designed to protect against SRAM errors that are introduced when the data resides within the host FIFO buffer. When the data reaches the head of the host FIFO buffer, the controller fetches the data from the buffer, uses the HFIFO parity to correct any errors, discards the HFIFO parity, and sends the data to the DRAM manager (❸). The DRAM manager buffers the data (which still contains the MPECC information) within DRAM (❹), and keeps track of the location of the buffered data inside the DRAM. When the controller is ready to write the data to the NAND flash memory, the DRAM manager reads the data from DRAM. Then, the controller uses the MPECC information to correct any errors, and discards the MPECC information. The controller then encodes the data into an ECC codeword (❺), generates CRC parity for the codeword, and then writes both the codeword and the CRC parity to a NAND flash FIFO buffer (❻) [161]. When the codeword reaches the head of this buffer, the controller uses CRC parity to correct any errors in the codeword, and then dispatches the data to the flash interface (❼), which writes the data to the NAND flash memory. The read data path of the controller (the green dotted line shown in Figure 3) performs the same procedure as the write data path, but in reverse order [161].

Aside from buffering data along the write and read paths, the controller uses the DRAM to store essential metadata, such as the table that maps each host data address to a physical block address within the NAND flash memory [122, 161]. In the metadata path of the controller (the blue dashed line shown in Figure 3), the metadata is often read from or written to DRAM by the firmware processors. In order to ensure correct operation of the SSD, the metadata must not contain any errors. As a result, the controller uses memory protection ECC (MPECC) for the metadata stored within DRAM [118, 161], just as it did to buffer data along the write and read data paths. Due to the lower rate of errors in DRAM compared to NAND flash memory (see Section 8), the employed memory protection ECC algorithms are not as strong as BCH or LDPC. We describe common ECC algorithms employed for DRAM error correction in Section 8.

**Bad Block Management.** Due to process variation or uneven wearout, a small number of flash blocks may have a much higher raw bit error rate (RBER) than an average flash block. Mitigating or tolerating the RBER on these flash blocks often requires a much higher cost than the benefit of using them. Thus, it is more efficient to identify and record these blocks as *bad blocks*, and avoid using them to store useful data. There are two types of bad blocks: *original bad blocks* (OBBs), which are defective due to manufacturing issues (e.g., process variation), and *growth bad blocks* (GBBs), which fail during runtime [179].

The flash vendor performs extensive testing, known as *bad block scanning*, to identify OBBs when a flash chip is manufactured [125]. Initially, all blocks are kept in the erased state, and contain the value 0xFF in each byte (see Section 3.1). Inside each OBB, the bad block scanning procedure writes a specific data value (e.g., 0x00) to a specific byte location within the block that indicates the block status. A good block (i.e., a block without defects) is not modified, and thus its block status byte remains at the value 0xFF. When the SSD is powered up for the first time, the SSD controller iterates through all blocks and checks the value stored in the block status byte of each block. Any block that does not contain the value 0xFF is marked as bad, and is recorded in a *bad block table* stored in the controller. A small number of blocks in each plane are set aside as *reserved blocks* (i.e., blocks that are not used during normal operation), and the bad block table automatically remaps any operation originally destined to an OBB to one of the reserved blocks. The bad block table remaps an OBB to a reserved block in the same plane, to ensure that the SSD maintains the same degree of parallelism when writing to a superpage, thus avoiding performance loss. Less than 2% of all blocks in the SSD are expected to be OBBs [146].

The SSD identifies growth bad blocks during runtime by monitoring the status of each block. Each superblock contains a bit vector indicating which of its blocks are GBBs. After each program or erase operation to a block, the SSD reads the *status reporting registers* to check the operation status. If the operation has failed, the controller marks the block as a GBB in the superblock bit vector. At this point, the controller uses superpage-level parity to recover the data that was stored in the GBB (see *Superpage-Level Parity* below), and *all data in the superblock* is copied to a different superblock. The superblock containing the GBB is then erased. When the superblock is subsequently opened, blocks marked as GBBs are *not* used, but the remaining blocks can store new data.

**Superpage-Level Parity.** In addition to ECC to protect against bit-level errors, many SSDs employ RAID-like parity [49, 78, 124, 155]. The key idea is to store parity information within each superpage to protect data from ECC failures that occur within a single chip or plane. Figure 4 shows an example of how the ECC and parity information are organized within a superpage. For a superpage that spans across multiple chips, dies, and planes, the pages stored within one die or one plane (depending on the implementation) are used to store parity information for the remaining pages. Without loss of generality, we assume for the rest of this section that a superpage that spans $c$ chips and $d$ dies per chip stores parity information in the pages of a single die (which we call the *parity die*), and that it stores user data in the pages of the remaining $(c \times d) - 1$ dies. When all of the user data is written to the superpage, the SSD controller XORs the data together one plane at a time (e.g., in Figure 4, all of the pages in Plane 0 are XORed with each other), which produces the parity data for that plane. This parity data is written to the corresponding plane in the parity die, e.g., Plane 0 page in Die $(c \times d) - 1$ in the figure.

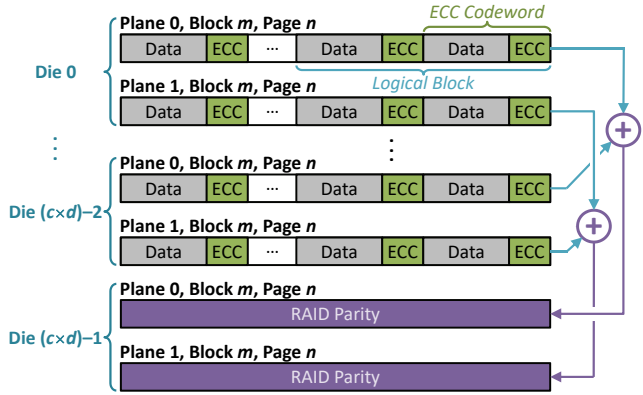

Figure 4: Example layout of ECC codewords, logical blocks, and superpage-level parity for superpage $n$ in superblock $m$. In this example, we assume that a logical block contains two codewords.

The SSD controller invokes superpage-level parity when an ECC failure occurs during a host software (e.g., OS, file system) access to the SSD. The host software accesses data at the granularity of a *logical block* (LB), which is indexed by a *logical block address* (LBA). Typically, an LB is 4 kB in size, and consists of several ECC codewords (which are usually 512 BB to 2 kB in size) stored consecutively within a flash memory page, as shown in Figure 4. During the LB access, a read failure can occur for one of two reasons. First, it is possible that the LB data is stored within a *hidden* GBB (i.e., a GBB that has not yet been detected and excluded by the bad block manager). The probability of storing data in a hidden GBB is quantified as $P_{HGBB}$. Note that because bad block management successfully identifies and excludes most GBBs, $P_{HGBB}$ is much lower than the total fraction of GBBs within an SSD. Second, it is possible that at least one ECC codeword within the LB has *failed* (i.e., the codeword contains an error that cannot be corrected by ECC). The probability that a codeword fails is $P_{ECFR}$ (see *Error-Correcting Codes* above). For an LB that contains $K$ ECC codewords, we can model $P_{LBFail}$, the overall probability that an LB access fails (i.e., the rate at which superpage-level parity needs to be invoked), as:

$$P_{LBFail} = P_{HGBB} + [1 - P_{HGBB}] \times [1 - (1 - P_{ECFR})^K] \quad (2)$$

In Equation 2, $P_{LBFail}$ consists of (1) the probability that an LB is inside a hidden GBB (left side of the addition); and (2) for an LB that is not in a hidden GBB, the probability of any codeword failing (right side of the addition).

When a read failure occurs for an LB in plane $p$, the SSD controller reconstructs the data using the other LBs in the same superpage.

To do this, the controller reads the LBs stored in plane $p$ in the other $(c \times d) - 1$ dies of the superpage, including the LBs in the parity die. The controller then XORs all of these LBs together, which retrieves the data that was originally stored in the LB whose access failed. In order to correctly recover the failed data, all of the LBs from the $(c \times d) - 1$ dies must be correctly read. The overall superpage-level parity failure probability $P_{parity}$ (i.e., the probability that more than one LB contains a failure) for an SSD with $c$ chips of flash memory, with $d$ dies per chip, can be modeled as [155]:

$$P_{parity} = P_{LBFail} \times [1 - (1 - P_{LBFail})^{(c \times d) - 1}] \quad (3)$$

Thus, by designating one of the dies to contain parity information (in a fashion similar to RAID 4 [155]), the SSD can tolerate the *complete failure* of the superpage data in one die without experiencing data loss during an LB access.

### 2.4 Design Tradeoffs for Reliability

Several design decisions impact the SSD *lifetime* (i.e., the duration of time that the SSD can be used within a bounded probability of error without exceeding a given performance overhead). To capture the tradeoff between these decisions and lifetime, SSD manufacturers use the following model:

$$\text{Lifetime (Years)} = \frac{\text{PEC} \times (1 + \text{OP})}{365 \times \text{DWPD} \times \text{WA} \times R_{compress}} \quad (4)$$

In Equation 4, the numerator is the total number of full drive writes the SSD can endure (i.e., for a drive with an $X$-byte capacity, the number of times $X$ bytes of data can be written). The number of full drive writes is calculated as the product of PEC, the total P/E cycle *endurance* of each flash block (i.e., the number of P/E cycles the block can sustain before its raw error rate exceeds the ECC correction capability), and $1 + \text{OP}$, where OP is the *overprovisioning factor* selected by the manufacturer. Manufacturers overprovision the flash drive by providing more physical block addresses, or PBAs, to the SSD controller than the *advertised capacity* of the drive, i.e., the number of logical block addresses (LBAs) available to the operating system. Overprovisioning improves performance and endurance, by providing additional free space in the SSD so that maintenance operations can take place without stalling host requests. OP is calculated as:

$$\text{OP} = \frac{\text{PBA count} - \text{LBA count}}{\text{LBA count}} \quad (5)$$

The denominator in Equation 4 is the number of full drive writes per year, which is calculated as the product of days per year (i.e., 365), DWPD, and the ratio between the total size of the data written to flash media and the size of the data sent by the host (i.e., $\text{WA} \times R_{compress}$). DWPD is the number of full disk writes per day (i.e., the number of times per day the OS writes the advertised capacity's worth of data). DWPD is typically less than 1 for read-intensive applications, and could be greater than 5 for write-intensive applications [24]. WA (*write amplification*) is the ratio between the amount of data written into NAND flash memory by the controller over the amount of data written by the host machine. Write amplification occurs because various procedures (e.g., garbage collection [35, 202]; and remapping-based refresh, Section 5.3) in the SSD perform additional writes in the background. For example, when garbage collection selects a block to erase, the pages that are remapped to a new block require background writes. $R_{compress}$, or the compression ratio, is the ratio between the size of the compressed data and the size of the uncompressed data, and is a function of the entropy of the stored data and the efficiency of the compression algorithms employed in the SSD controller. In Equation 4, DWPD and $R_{compress}$ are largely determined by the workload and data compressibility, and cannot be changed to optimize flash lifetime. For controllers that do not implement compression, we set R compress to 1. However, the SSD controller can trade off other parameters between one another to optimize flash lifetime. We discuss the most salient tradeoffs next.

**Tradeoff Between Write Amplification and Overprovisioning.** As mentioned in Section 2.3, due to the granularity mismatch between flash erase and program operations, garbage collection occasionally remaps remaining valid pages from a selected block to a new flash block, in order to avoid block-internal fragmentation. This remapping causes additional flash memory writes, leading to *write amplification*. In an SSD with more overprovisioned capacity, the amount of write amplification decreases, as the blocks selected for garbage collection are older and tend to have fewer valid pages. For a greedy garbage collection algorithm and a random-access workload, the correlation between WA and OP can be calculated [48, 67], as shown in Figure 5. In an ideal SSD, both WA and OP should be minimal, i.e., WA = 1 and OP = 0%, but in reality there is a tradeoff between these parameters: when one increases, the other decreases. As Figure 5 shows, WA can be reduced by increasing OP, and with an infinite amount of OP, WA converges to 1. However, the reduction of WA is smaller when OP is large, resulting in diminishing returns.

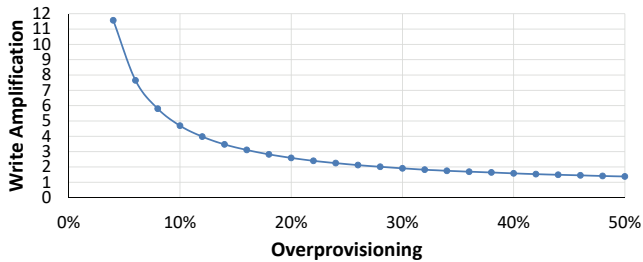

**Figure 5: Relationship between write amplification (WA) and the overprovisioning factor (OP).**

In reality, the relationship between WA and OP is also a function of the storage space utilization of the SSD. When the storage space is *not* fully utilized, many more pages are available, reducing the need to invoke garbage collection, and thus WA can approach 1 without the need for a large amount of OP.

**Tradeoff Between P/E Cycle Endurance and Overprovisioning.** PEC and OP can be traded against each other by adjusting the amount of redundancy used for error correction, such as ECC and superpage-level parity (as discussed in Section 2.3). As the error correction capability increases, PEC increases because the SSD can tolerate the higher raw bit error rate that occurs at a higher P/E cycle count. However, this comes at a cost of reducing the amount of space available for OP, since a stronger error correction capability requires higher redundancy (i.e., more space). Table 1 shows the corresponding OP for four different error correction configurations for an example SSD with 2.0 TB of advertised capacity and 2.4 TB (20% extra) of physical space. In this table, the top two configurations use ECC-1 with a coding rate of 0.93, and the bottom two configurations use ECC-2 with a coding rate of 0.90, which has higher redundancy than ECC-1. Thus, the ECC-2 configurations have a lower OP than the top two. ECC-2, with its higher redundancy, can correct a greater number of raw bit errors, which in turn increases the P/E cycle endurance of the SSD. Similarly, the two configurations with superpage-level parity have a lower OP than configurations without superpage-level parity, as parity uses a portion of the overprovisioned space to store the parity bits.

**Table 1: Tradeoff between strength of error correction configuration and amount of SSD space left for overprovisioning.**

| Error Correction Configuration | Overprovisioning Factor |
|---|---|
| ECC-1 (0.93), no superpage-level parity | 11.6% |
| ECC-1 (0.93), with superpage-level parity | 8.1% |
| ECC-2 (0.90), no superpage-level parity | 8.0% |
| ECC-2 (0.90), with superpage-level parity | 4.6% |

When the ECC correction strength is increased, the amount of overprovisioning in the SSD decreases, which in turn increases the amount of write amplification that takes place. Manufacturers must find and use the correct tradeoff between ECC correction strength and the overprovisioning factor, based on which of the two is expected to provide greater reliability for the target applications of the SSD.

## 3 NAND FLASH MEMORY BASICS

A number of underlying properties of the NAND flash memory used within the SSD affect SSD management, performance, and reliability [5, 8, 126]. In this section, we present a primer on NAND flash memory and its operation, to prepare the reader for understanding our further discussion on error sources (Section 4) and mitigation mechanisms (Section 5). Recall from Section 2.1 that within each plane, flash cells are organized as multiple 2D arrays known as flash blocks, each of which contains multiple pages of data, where a page is the granularity at which the host reads and writes data. We first discuss how data is stored in NAND flash memory. We then introduce the three basic operations supported by NAND flash memory: read, program, and erase.

### 3.1 Storing Data in a Flash Cell

NAND flash memory stores data as the *threshold voltage* of each flash cell, which is made up of a *floating gate transistor*. Figure 6 shows a cross section of a floating gate transistor. On top of a flash cell is the *control gate* (CG) and below is the floating gate (FG). The floating gate is insulated on both sides, on top by an interpoly oxide layer and at the bottom by a tunnel oxide layer. As a result, the electrons programmed on the floating gate do not discharge even when flash memory is powered off.

For *single-level cell* (SLC) NAND flash, each flash cell stores a 1-bit value, and can be programmed to one of two threshold voltage states, which we call the ER and P1 states. *Multi-level cell* (MLC) NAND flash stores a 2-bit value in each cell, with four possible states (ER, P1, P2, and P3), and *triple-level cell* (TLC) NAND flash stores a 3-bit value in each cell with eight possible states (ER, P1–P7). Each state represents a different value, and is assigned a *voltage window* within the range of all possible threshold voltages. Due to variation across *program* operations, the threshold voltage of flash cells programmed to the same state is initially distributed across this voltage window.

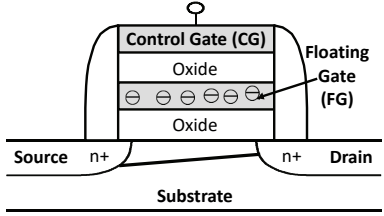

Figure 6: Flash cell (i.e., floating gate transistor) cross section.

Figure 7 illustrates the threshold voltage distribution of MLC (top) and TLC (bottom) NAND flash memories. The x-axis shows the threshold voltage ($V_{th}$), which spans a certain voltage range. The y-axis shows the probability density of each voltage level across all flash memory cells. The threshold voltage distribution of each threshold voltage state can be represented as a probability density curve that spans over the state's voltage window.

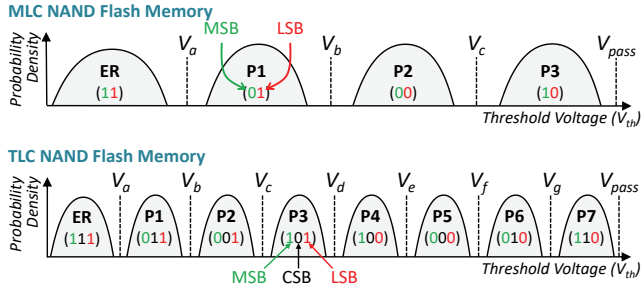

Figure 7: Threshold voltage distribution of MLC (top) and TLC (bottom) NAND flash memory.

We label the distribution curve for each state with the name of the state and a corresponding bit value. Note that some manufacturers may choose to use a different mapping of values to different states. The bit values of adjacent states are separated by a Hamming distance of 1. We break down the bit values for MLC into the most significant bit (MSB) and least significant bit (LSB), while TLC is broken down into the MSB, the center significant bit (CSB), and the LSB. The boundaries between neighboring threshold voltage windows, which are labeled as $V_a$, $V_b$, and $V_c$ for the MLC distribution in Figure 7, are referred to as *read reference voltages*. These voltages are used by the SSD controller to identify the voltage window (i.e., state) of each cell upon reading the cell.

## 3.2 Flash Block Design

Figure 8 shows the high-level internal organization of a NAND flash memory block. Each block contains multiple rows of cells (typically 128–512 rows). Each row of cells is connected together by a common *wordline* (WL, shown horizontally in Figure 8), typically spanning 32K–64K cells. All of the cells along the wordline are logically combined to form a page in an SLC NAND flash memory. For an MLC NAND flash memory, the MSBs of all cells on the same wordline are combined to form an *MSB page*, and the LSBs of all cells on the wordline are combined to form an *LSB page*. Similarly, a TLC NAND flash memory logically combines the MSBs on each wordline to form an MSB page, the CSBs on each wordline to form a *CSB page*, and the LSBs on each wordline to form an LSB page. In MLC NAND flash memory, each flash block contains 256–1024 flash pages, each of which are typically 8–16 kB in size.

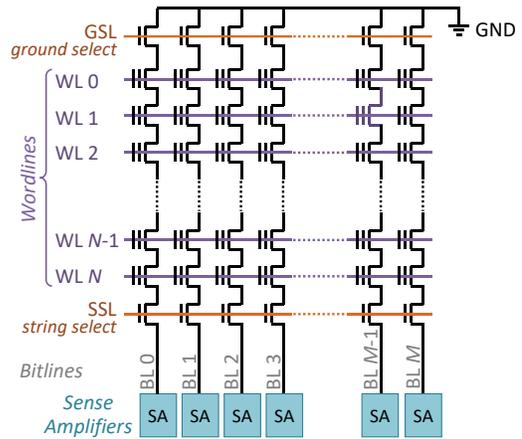

Figure 8: Internal organization of a flash block.

Within a block, all cells in the same column are connected in series to form a *bitline* (BL, shown vertically in Figure 8) or *string*. All cells in a bitline share a common ground (GND) on one end, and a common *sense amplifier* (SA) on the other for reading the threshold voltage of one of the cells when decoding data. Bitline operations are controlled by turning the *ground select line* (GSL) and *string select line* (SSL) transistor of each bitline on or off. The SSL transistor is used to enable operations on a bitline, and the GSL transistor is used to connect the bitline to ground during a read operation [127]. The use of a common bitline across multiple rows reduces the amount of circuit area required for read and write operations to a block, improving storage density.

## 3.3 Read Operation

Data can be read from NAND flash memory by applying read reference voltages onto the control gate of each cell, to sense the cell's threshold voltage. To read the value stored in a single-level cell, we need to distinguish only the state with a bit value of 1 from the state with a bit value of 0. This requires us to use only a single read reference voltage. Likewise, to read the LSB of a multi-level cell, we need to distinguish only the states where the LSB value is 1 (ER and P1) from the states where the LSB value is 0 (P2 and P3), which we can do with a single read reference voltage ($V_b$ in the top half of Figure 7). To read the MSB page, we need to distinguish the states with an MSB value of 1 (ER and P3) from those with an MSB value of 0 (P1 and P2). Therefore, we need to determine whether the threshold voltage of the cell falls between $V_a$ and $V_c$, requiring us to apply each of these two read reference voltages (which can require up to two consecutive read operations) to determine the MSB.

Reading data from a triple-level cell is similar to the data read procedure for a multi-level cell. Reading the LSB for TLC again requires applying only a single read reference voltage ($V_d$ in the bottom half of Figure 7). Reading the CSB requires two read reference voltages to be applied, and reading the MSB requires four read reference voltages to be applied.

As Figure 8 shows, cells from multiple wordlines (WL in the figure) are connected in series on a *shared* bitline (BL) to the sense amplifier, which drives the value that is being read from the block onto the memory channel for the plane. In order to read from a single cell on the bitline, *all of the other cells* (i.e., *unread* cells) on the same bitline must be switched on to allow the value that is being read to propagate through to the sense amplifier. The NAND flash memory achieves this by applying the *pass-through voltage* onto the wordlines of the unread cells, as shown in Figure 9a. When the pass-through voltage (i.e., the maximum possible threshold voltage $V_{pass}$) is applied to a flash cell, the source and the drain of the cell transistor are connected, regardless of the voltage of the floating gate. Modern flash memories guarantee that all *unread* cells are *passed through* to minimize errors during the read operation [16].

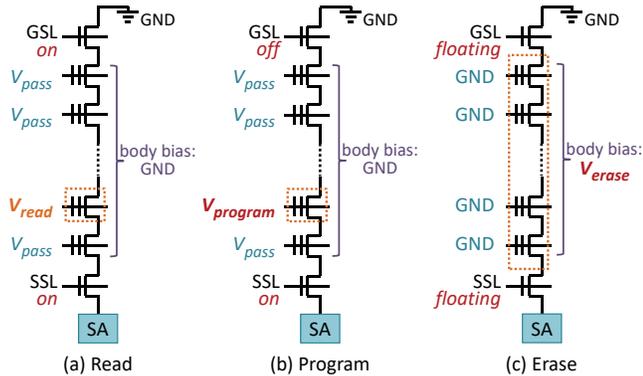

Figure 9: Voltages applied to flash cell transistors on a bitline to perform (a) read, (b) program, and (c) erase operations.

### 3.4 Program and Erase Operations

The threshold voltage of a floating gate transistor is controlled through the injection and ejection of electrons through the tunnel oxide of the transistor, which is enabled by the Fowler–Nordheim (FN) tunneling effect [5, 52, 156]. The tunneling current ($J_{FN}$) [8, 156] can be modeled as:

$$J_{FN} = \alpha_{FN} E_{ox}^2 e^{-\beta_{FN}/E_{ox}} \qquad (6)$$

In Equation 6, $\alpha_{FN}$ and $\beta_{FN}$ are constants, and $E_{ox}$ is the electric field strength in the tunnel oxide. As Equation 6 shows, $J_{FN}$ is exponentially correlated with $E_{ox}$.

During a program operation, electrons are injected into the floating gate of the flash cell from the substrate when applying a high positive voltage to the control gate (see Figure 6 for a diagram of the flash cell). The pass-through voltage is applied to all of the other cells on the same bitline as the cell that is being programmed as shown in Figure 9b. When data is programmed, charge is transferred into the floating gate through FN tunneling by repeatedly pulsing the programming voltage, in a procedure known as *incremental step-pulse programming* (ISPP) [5, 126, 175, 185]. During ISPP, a high programming voltage ($V_{program}$) is applied for a very short period, which we refer to as a *step-pulse*. ISPP then verifies the current voltage of the cell using the voltage $V_{verify}$. ISPP repeats the process of applying a step-pulse and verifying the voltage until the cell reaches the desired target voltage. In the modern all-bitline NAND flash memory, all flash cells in a single wordline are programmed concurrently. During programming, when a cell along the wordline reaches its target voltage but other cells have yet to reach their target voltage, ISPP *inhibits* programming pulses to the cell by turning off the SSL transistor of the cell's bitline.

In SLC NAND flash and older MLC NAND flash, *one-shot programming* is used, where all of the ISPP step-pulses required to program a cell are applied back to back until all cells in the wordline are fully programmed. One-shot programming does *not* interleave the program operations to a wordline with the program operations to another wordline. In newer MLC NAND flash, the lack of interleaving between program operations can introduce a significant amount of cell-to-cell program interference on the cells of immediately-adjacent wordlines (see Section 4.3).

To reduce the impact of program interference, the controller employs *two-step programming* for sub-40 nm MLC NAND flash [18, 151]: it first programs the LSBs into the erased cells of an unprogrammed wordline, and then programs the MSBs of the cells using a separate program operation [12, 15, 149, 151]. Between the programming of the LSBs and the MSBs, the controller programs the LSBs of the cells in the wordline immediately above [12, 15, 149, 151]. Figure 10 illustrates the two-step programming algorithm. In the first step, a flash cell is *partially programmed* based on its LSB value, either staying in the ER state if the LSB value is 1, or moving to a temporary state (TP) if the LSB value is 0. The TP state has a mean voltage that falls between states P1 and P2. In the second step, the LSB data is first read back into an internal buffer register within the flash chip to determine the cell's current threshold voltage state, and then further programming pulses are applied based on the MSB data to increase the cell's threshold voltage to fall within the voltage window of its final state. Programming in MLC NAND flash is discussed in detail in [12] and [15].

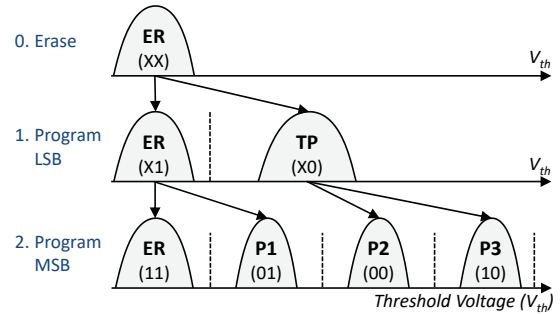

Figure 10: Two-step programming algorithm for MLC flash.

TLC NAND flash takes a similar approach to the two-step programming of MLC, with a mechanism known as *foggy-fine programming* [111], which is illustrated in Figure 11. The flash cell is first partially programmed based on its LSB value, using a *binary* programming step in which very large ISPP step-pulses are used to significantly increase the voltage level. Then, the flash cell is partially programmed again based on its CSB and MSB values to a new set of temporary states (these steps are referred to as *foggy* programming, which uses smaller ISPP step-pulses than binary programming). Due to the higher potential for errors during TLC programming as a result of the narrower voltage windows, all of the programmed bit values are buffered after the binary and foggy programming steps into SLC buffers that are reserved in each chip/plane. Finally, *fine* programming takes place, where these bit values are read from the SLC buffers, and the smallest ISPP step-pulses are applied to set each cell to its final threshold voltage state. The purpose of this last fine programming step is to fine tune the threshold voltage such that the threshold voltage distributions are tightened (bottom of Figure 11).

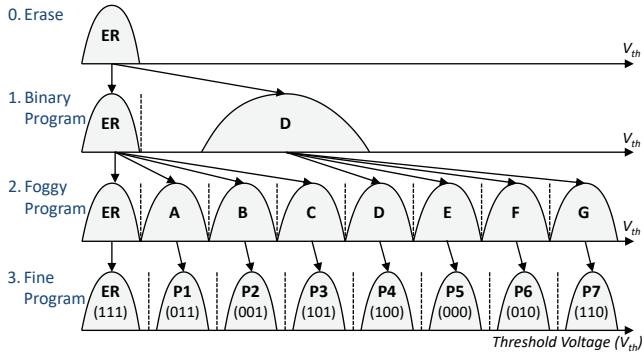

Figure 11: Foggy-fine programming algorithm for TLC flash.

Though programming sets a flash cell to a specific threshold voltage using programming pulses, the voltage of the cell can drift over time after programming. When no external voltage is applied to any of the electrodes (i.e., CG, source, and drain) of a flash cell, an electric field still exists between the FG and the substrate, generated by the charge present in the FG. This is called the *intrinsic electric field* [8], and it generates *stress-induced leakage current* (SILC) [5, 46, 143], a weak tunneling current that leaks charge away from the FG. As a result, the voltage that a cell is programmed to may not be the same as the voltage read for that cell at a subsequent time.

In NAND flash, a cell can be reprogrammed with new data *only after* the existing data in the cell is erased. This is because ISPP can only *increase* the voltage of the cell. The erase operation resets the threshold voltage state of *all cells in the flash block* to the ER state. During an erase operation, electrons are ejected from the FG of the flash cell into the substrate by inducing a high negative voltage on the cell transistor. The negative voltage is induced by setting the CG of the transistor to GND, and biasing the transistor body (i.e., the substrate) to a high voltage ($V_{erase}$), as shown in Figure 9c. Because all cells in a flash block share a common transistor substrate (i.e., the bodies of all transistors in the block are connected together), a flash block must be erased in its entirety [127].

## 4 NAND FLASH ERROR CHARACTERIZATION

Each block in NAND flash memory is used in a cyclic fashion, as is illustrated by the observed raw bit error rates seen over the lifetime of a flash memory block in Figure 12. At the beginning of a *cycle*, known as a *program/erase (P/E) cycle*, an erased block is *opened* (i.e., selected for programming). Data is then programmed into the open block one page at a time. After all of the pages are programmed, the block is closed, and none of the pages can be reprogrammed until the whole block is erased. At any point before erasing, read operations can be performed on a *valid* programmed page (i.e., a page containing data that has not been modified by the host). A page is marked as invalid when the data stored at that page's logical address by the host is modified. As ISPP can only inject more charge into the floating gate but cannot remove charge from the gate, it is not possible to modify data to a new arbitrary value *in place* within existing NAND flash memories. Once the block is erased, the P/E cycling behavior repeats until the block is *worn out* (i.e., the block can no longer avoid data loss over the course of the minimum data retention period guaranteed by the manufacturer). Although the 5x-nm (i.e., 50–59 nm) generation of MLC NAND flash could endure ~10,000 P/E cycles per block before being worn out, modern 1x-nm (i.e., 15–19 nm) MLC and TLC NAND flash can endure only ~3,000 and ~1,000 P/E cycles per block, respectively [95, 120, 153, 206].

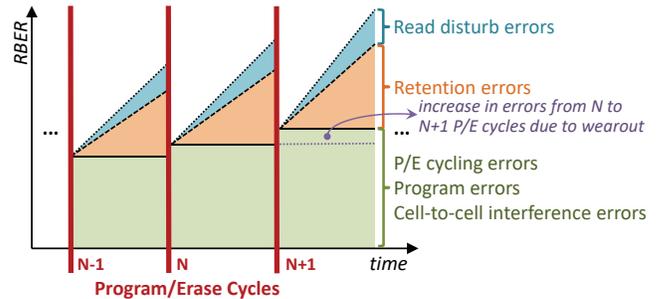

Figure 12: Pictorial depiction of errors accumulating within a NAND flash block as P/E cycle count increases.

As shown in Figure 12, several different types of errors can be introduced at any point during the P/E cycling process: *P/E cycling errors*, *program errors*, errors due to *cell-to-cell program interference*, *data retention errors*, and errors due to *read disturb*. As discussed in Section 3.1, the threshold voltage of flash cells programmed to the same state is distributed across a voltage window due to variation across program operations and across different flash cells. Several types of errors introduced during the P/E cycling process, such as data retention and read disturb, cause the threshold voltage distribution of each state to shift and widen. Due to the shift and widening, the tails of the distributions of each state can enter the margin that originally existed between each of the two neighboring states' distributions. Thus, the threshold voltage distributions of different states can start overlapping, as shown in Figure 13. When the distributions overlap with each other, the read reference voltages can no longer correctly identify the state of some flash cells in the overlapping region, leading to *raw bit errors* during a read operation.

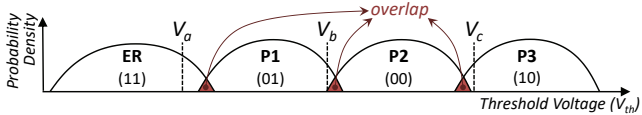

Figure 13: Threshold voltage distribution shifts and widening can cause the distributions of two neighboring states to overlap with each other (compare to Figure 7), leading to read errors.

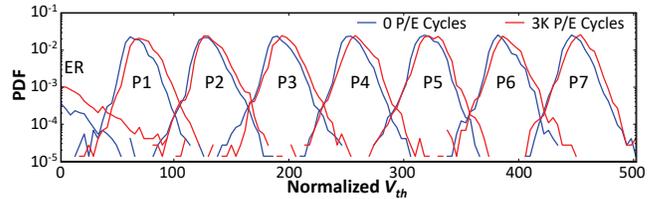

Figure 14: Threshold voltage distribution of TLC NAND flash memory after 0 P/E cycles and 3,000 P/E cycles.

In this section, we discuss the causes of each type of error in detail, and characterize the impact that each error type has on the amount of raw bit errors occurring within NAND flash memory. We use an FPGA-based testing platform [13] to characterize state-of-the-art TLC NAND flash chips. We use the read-retry operation present in NAND flash devices to accurately read the cell threshold voltage [15–18, 24, 26, 53, 116, 150] (for a detailed description of the read-retry operation, see Section 5.4). As absolute threshold voltage values are proprietary information to flash vendors, we present our results using normalized voltages, where the nominal maximum value of $V_{th}$ is equal to 512 in our normalized scale, and where 0 represents GND. We also describe characterization results and observations for MLC NAND flash chips. These MLC NAND results are taken from our prior works [10, 12, 14–18, 24–26, 116], which provide more detailed error characterization results and analyses. To our knowledge, this paper provides the first experimental characterization and analysis of errors in real *TLC* NAND flash memory chips.

We later discuss mitigation techniques for these flash memory errors in Section 5, and provide procedures to recover in the event of data loss in Section 6.

### 4.1 P/E Cycling Errors

A P/E cycling error occurs when either (1) an erase operation fails to reset a cell to the ER state; or (2) when a program operation fails to set the cell to the desired target state. P/E cycling errors occur because electrons become trapped in the tunnel oxide after stress from repeated P/E cycles. Errors due to such electron trapping (which we refer to as *P/E cycling noise*) continue to accumulate over the lifetime of a NAND flash block. This behavior is called *wearout*, and it refers to the phenomenon where, as more writes are performed to a block, there are a greater number of raw bit errors that must be corrected, exhausting more of the fixed error correction capability of the ECC (see Section 2.3).

Figure 14 shows the threshold voltage distribution of TLC NAND flash memory after 0 P/E cycles and after 3,000 P/E cycles, without any retention or read disturb errors present (which we ensure by reading the data *immediately* after programming). The mean and standard deviation of each state's distribution are provided in Table 4 in the Appendix (for other P/E cycle counts as well). We make two observations from the two distributions. First, as the P/E cycle count increases, each state's threshold voltage distribution systematically (1) shifts to the right and (2) becomes wider. Second, the amount of the shift is greater for lower-voltage states (e.g., the ER and P1 states) than it is for higher-voltage states (e.g., the P7 state).

The threshold voltage distribution shift occurs because as more P/E cycles take place, the quality of the tunnel oxide degrades, allowing electrons to tunnel through the oxide more easily [129]. As a result, if the same ISPP conditions (e.g., programming voltage, step-pulse size, program time) are applied throughout the lifetime of the NAND flash memory, more electrons are injected during programming as a flash memory block wears out, leading to higher threshold voltages, i.e., the right shift of the distribution. The distribution of each state widens due to the process variation present in (1) the wearout process, and (2) the cell's structural characteristics. As the distribution of each voltage state widens, more overlap occurs between neighboring distributions, making it less likely for a read reference voltage to determine the correct value of the cells in the overlapping regions, which leads to a greater number of raw bit errors.

The threshold voltage distribution trends we observe here for TLC NAND flash memory trends are similar to trends observed previously for MLC NAND flash memory [14, 15, 116, 153], although the MLC NAND flash characterizations reported in past studies span up to a larger P/E cycle count than the TLC experiments due to the greater endurance of MLC NAND flash memory. More findings on the nature of wearout and the impact of wearout on NAND flash memory errors and lifetime can be found in our prior work [10, 14, 15, 116].

### 4.2 Program Errors

Program errors occur when data read directly from the NAND flash array contains errors, and the erroneous values are used to program the new data. Program errors occur in two major cases: (1) partial programming during two-step or foggy-fine programming, and (2) *copyback* (i.e., when data is copied inside the NAND flash memory during a maintenance operation) [68]. During two-step programming for MLC NAND flash memory (see Figure 10), in between the LSB and MSB programming steps of a cell, threshold voltage shifts can occur on the partially-programmed cell. These shifts occur because several other read and program operations to cells in *other* pages within the same block may take place, causing interference to the partially-programmed cell. Figure 15 illustrates how the threshold distribution of the ER state widens and shifts to the right after the LSB value is programmed (step 1 in the figure). The widening and shifting of the distribution causes some cells that were originally partially programmed to the ER state (with an LSB value of 1) to be misread as being in the TP state (with an LSB value of 0) during the *second* programming step (step 2 in the figure). As shown in Figure 15, the misread LSB value leads to a program error when the final cell threshold voltage is programmed [12, 116, 153]. Some cells that should have been programmed to the P1 state (representing the value 01) are instead programmed to the P2 state (with the value 00), and some cells that should have been programmed to the ER state (representing the value 11) are instead programmed to the P3 state (with the value 10).

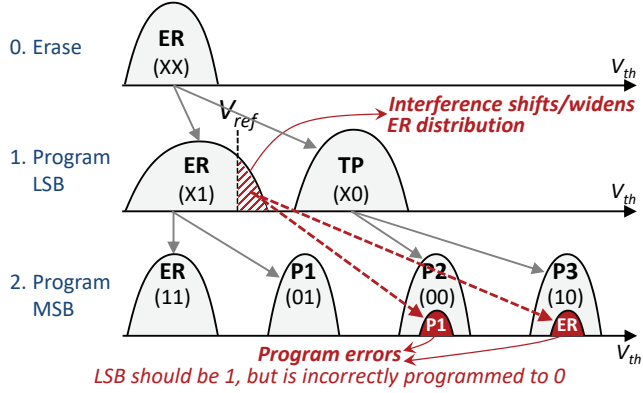

Figure 15: Impact of program errors during two-step programming on cell threshold voltage distribution.

The incorrect values that are read before the second programming step are *not* corrected by ECC, as they are read directly inside the NAND flash array, without involving the controller (where the ECC engine resides). Similarly, during foggy-fine programming for TLC NAND flash (see Figure 11), the data may be read incorrectly from the SLC buffers used to store the contents of partially-programmed wordlines, leading to errors during the fine programming step. Program errors occur during *copyback* [68] when valid data is read out from a block during maintenance operations (e.g., a block about to be garbage collected) and reprogrammed into a new block, as copyback operations do *not* go through the SSD controller.

Program errors that occur during partial programming predominantly shift data from lower-voltage states to higher-voltage states. For example, in MLC NAND flash, program errors predominantly shift data that should be in the ER state (11) into the P3 state (10), or data that should be in the P1 state (01) into the P2 state (00) [12]. This occurs because MSB programming can only *increase* (and not reduce) the threshold voltage of the cell from its partially-programmed voltage (and thus cannot move a multi-level cell that should be in the P3 state into the ER state, or one that should be in the P2 state into the P1 state). TLC NAND flash is much less susceptible to program errors than MLC NAND flash, as the data read from the SLC buffers in TLC NAND flash has a much lower error rate than data read from a partially-programmed MLC NAND flash wordline [167].

From a rigorous experimental characterization of modern MLC NAND flash memory chips [12], we find that program errors occur primarily due to two types of errors affecting the partially-programmed data. First, cell-to-cell program interference (Section 4.3) on a partially-programmed wordline is no longer negligible in newer NAND flash memory compared to older NAND flash memory, due to manufacturing process scaling. As flash cells become smaller and are placed closer to each other, cells in partially-programmed wordlines become more susceptible to bit flips. Second, partially-programmed cells are more susceptible to read disturb errors than fully-programmed cells (Section 4.5), as the threshold voltages stored in these cells are no more than approximately half of $V_{pass}$ [12], and cells with lower threshold voltages are more likely to experience read disturb errors.

More findings on the nature of program errors and the impact of program errors on NAND flash memory lifetime can be found in our prior work [12, 116].

### 4.3 Cell-to-Cell Program Interference Errors

Program interference refers to the phenomenon where the programming of a flash cell induces errors on adjacent flash cells within a flash block [18, 26, 44, 56, 108]. The interference occurs due to *parasitic capacitance coupling* between these cells. As a result, when the threshold voltage of an adjacent flash cell increases, the threshold voltage of the *victim cell* increases as well. The unintended threshold voltage shifts can eventually move a cell into a different state than the one it was originally programmed to, leading to a bit error.

We have shown, based on our experimental analysis of modern MLC NAND flash memory chips, that the threshold voltage change of the victim cell can be accurately modeled as a linear combination of the threshold voltage changes of the adjacent cells when they are programmed, using linear regression with least-square-error estimation [18, 26]. The cells that are physically located immediately next to the victim cell (called the *immediately-adjacent cells*) are the major contributors to the cell-to-cell interference of a victim cell [18]. Figure 16 shows the eight immediately-adjacent cells for a victim cell in 2D planar NAND flash memory.

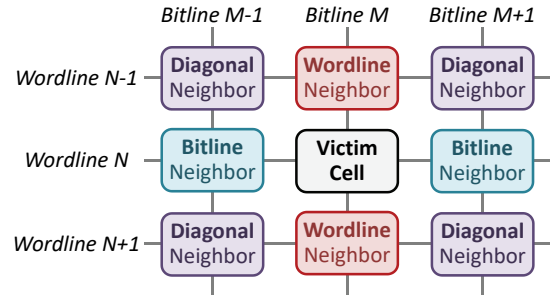

Figure 16: Immediately-adjacent cells that can induce program interference on a victim cell that is on wordline *N* and bitline *M*.

The amount of interference that program operations to the immediately-adjacent cells can induce on the victim cell is expressed as:

$$\Delta V_{victim} = \sum_X K_X \Delta V_X \qquad (7)$$

where $\Delta V_{victim}$ is the change in voltage of the victim cell due to cell-to-cell program interference, $K_X$ is the *coupling coefficient* between cell $X$ and the victim cell, and $\Delta V_X$ is the threshold voltage change of cell $X$ during programming. Table 2 lists the coupling coefficients for both 2y-nm and 1x-nm NAND flash memory. We make two key observations from Table 2. First, we observe that the coupling coefficient is greatest for wordline neighbors (i.e., immediately-adjacent cells on the same bitline, but on a neighboring wordline) [18]. The coupling coefficient is directly related to the effective capacitance $C$ between cell $X$ and the victim cell, which can be calculated as:

$$C = \varepsilon S/d \quad (8)$$

where $\varepsilon$ is the permittivity, $S$ is the effective cell area of cell $X$ that faces the victim cell, and $d$ is the distance between the cells. Of the immediately-adjacent cells, the wordline neighbor cells have the greatest coupling capacitance with the victim cell, as they likely have a large effective facing area to, and a small distance from, the victim cell compared to other surrounding cells. Second, we observe that the coupling coefficient grows as the feature size decreases [18, 26]. As NAND flash memory process technology scales down to smaller feature sizes, cells become smaller and get closer to each other, which increases the effective capacitance between them. As a result, at smaller feature sizes, it is easier for an immediately-adjacent cell to induce program interference on a victim cell. We conclude that (1) the program interference an immediately-adjacent cell induces on a victim cell is primarily determined by the distance between the cells and the immediately-adjacent cell's effective area facing the victim cell; and (2) the wordline neighbor cell causes the highest such interference, based on empirical measurements.

**Table 2: Coupling coefficients for immediately-adjacent cells.**

| Process Technology | Wordline Neighbor | Bitline Neighbor | Diagonal Neighbor |
|---|---|---|---|
| **2y-nm** | 0.060 | 0.032 | 0.012 |
| **1x-nm** | 0.110 | 0.055 | 0.020 |

Due to the order of program operations performed in NAND flash memory, many immediately-adjacent cells do *not* end up inducing interference after a victim cell is fully programmed (i.e., once the victim cell is at its target voltage). In modern all-bitline NAND flash memory, all flash cells on the same wordline are programmed at the same time, and wordlines are fully programmed sequentially (i.e., the cells on wordline $i$ are fully programmed before the cells on wordline $i + 1$). As a result, an immediately-adjacent cell on the wordline below the victim cell or on the same wordline as the victim cell does *not* induce program interference on a fully-programmed victim cell. Therefore, the major source of program interference on a fully-programmed victim cell is the programming of the wordline immediately above it.

Figure 17 shows how the threshold voltage distribution of a victim cell shifts when different values are programmed onto its immediately-adjacent cells in the wordline above the victim cell for MLC NAND flash, when one-shot programming is used. The amount by which the victim cell distribution shifts is directly correlated with the number of programming step-pulses applied to the immediately-adjacent cell. That is, when an immediately-adjacent cell is programmed to a higher-voltage state (which requires more step-pulses for programming), the victim cell distribution shifts further to the right [18]. When an immediately-adjacent cell is set to the ER state, no step-pulses are applied, as an unprogrammed cell is already in the ER state. Thus, no interference takes place. Note that the amount by which a fully-programmed victim cell distribution shifts is different when two-step programming is used, as a fully-programmed cell experiences interference from only one of the two programming steps of a neighboring wordline [12].

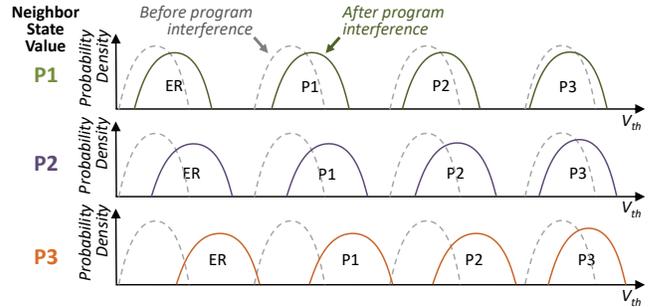

**Figure 17: Impact of cell-to-cell program interference on a victim cell during one-shot programming, depending on the value its neighboring cell is programmed to.**

More findings on the nature of cell-to-cell program interference and the impact of cell-to-cell program interference on NAND flash memory errors and lifetime can be found in our prior work [10, 12, 18, 26].

### 4.4 Data Retention Errors

Retention errors are caused by charge leakage over time after a flash cell is programmed, and are the dominant source of flash memory errors, as demonstrated previously [14, 17, 24, 25, 126, 177]. As flash memory process technology scales to smaller feature sizes, the capacitance of a flash cell, and the number of electrons stored on it, decreases. State-of-the-art (i.e., 1x-nm) MLC flash memory cells can store only ~100 electrons [206]. Gaining or losing several electrons on a cell can significantly change the cell's voltage level and eventually alter its state. Charge leakage is caused by the unavoidable trapping of charge in the tunnel oxide [17, 107]. The amount of trapped charge increases with the electrical stress induced by repeated program and erase operations, which degrade the insulating property of the oxide.

Two failure mechanisms of the tunnel oxide lead to retention loss. *Trap-assisted tunneling* (TAT) occurs because the trapped charge forms an electrical tunnel, which exacerbates the weak tunneling current, SILC (see Section 3.4). As a result of this TAT effect, the electrons present in the floating gate (FG) leak away much faster through the intrinsic electric field. Hence, the threshold voltage of the flash cell decreases over time. As the flash cell wears out with increasing P/E cycles, the amount of trapped charge also increases [17, 107], and so does the TAT effect. At high P/E cycles, the amount of trapped charge is large enough to form percolation paths that significantly hamper the insulating properties of the gate dielectric [17, 46], resulting in retention failure. *Charge detrapping*, where charge previously trapped in the tunnel oxide is freed spontaneously, can also occur over time [17, 46, 107, 200]. The charge polarity can be either negative (i.e., electrons) or positive (i.e., holes). Hence, charge detrapping can either decrease or increase the threshold voltage of a flash cell, depending on the polarity of the detrapped charge.

Figure 18 illustrates how the voltage distribution shifts for data we program into TLC NAND flash, as the data sits untouched over a period of one day, one month, and one year. The mean and standard deviation are provided in Table 5 in the Appendix (which includes data for other retention ages as well). These results are obtained from real flash memory chips we tested. We distill three major findings from these results, which are similar to our previously reported findings for retention behavior on MLC NAND flash memory [17].

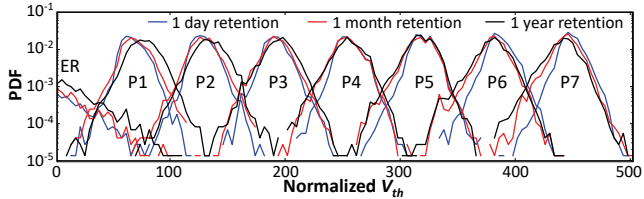

**Figure 18: Threshold voltage distribution for TLC NAND flash memory after one day, one month, and one year of retention time.**

First, as the *retention age* (i.e., the length of time after programming) of the data increases, the threshold voltage distributions of the higher-voltage states shift to lower voltages, while the threshold voltage distributions of the lower-voltage states shift to higher voltages. As the intrinsic electric field strength is higher for the cells in higher-voltage states, TAT is the dominant failure mechanism for these cells, which can *only* decrease the threshold voltage, as the resulting SILC can flow only in the direction of the intrinsic electric field generated by the electrons in the FG. Cells at the lowest-voltage states, where the intrinsic electric field strength is low, do not experience high TAT, and instead contain many *holes* (i.e., positive charge) that leak away as the retention age grows, leading to increase in threshold voltage.

Second, the threshold voltage distribution of each state becomes wider with retention age. Charge detrapping can cause cells to shift in either direction (i.e., toward lower or higher voltages), contributing to the widening of the distribution. The rate at which TAT occurs can also vary from cell to cell, as a result of process variation, which further widens the distribution.

Third, the threshold voltage distributions of higher-voltage states shift by a larger amount than the distributions of lower-voltage states. This is again a result of TAT. Cells at higher-voltage states have greater intrinsic electric field intensity, which leads to larger SILC. A cell where the SILC is larger experiences a greater drop in its threshold voltage than a cell where the SILC is smaller.

More findings on the nature of data retention and the impact of data retention behavior on NAND flash memory errors and lifetime can be found in our prior work [10, 14, 17, 24, 25].

### 4.5 Read Disturb Errors

Read disturb is a phenomenon in NAND flash memory where reading data from a flash cell can cause the threshold voltages of other (unread) cells in the same block to shift to a higher value [14, 16, 44, 56, 126, 148, 176]. While a single threshold voltage shift is small, such shifts can accumulate over time, eventually becoming large enough to alter the state of some cells and hence generate *read disturb errors*.

The failure mechanism of a read disturb error is similar to the mechanism of a normal program operation. A program operation applies a high programming voltage (e.g., +15 V) to the cell to change the cell's threshold voltage to the desired range. Similarly, a read operation applies a *high pass-through voltage* (e.g., +6 V) to *all other cells* that share the same bitline with the cell that is being read. Although the pass-through voltage is not as high as the programming voltage, it still generates a *weak programming effect* on the cells it is applied to [16], which can unintentionally change these cells' threshold voltages.

Figure 19 shows how read disturb errors impact threshold voltage distributions in real TLC NAND flash memory chips. We use blocks that have endured 2,000 P/E cycles, and we experimentally study the impact of read disturb on a single wordline in each block. We then read from a second wordline in the same block 1, 10K, and 100K times to induce different levels of read disturb. The mean and standard deviation of each distribution are provided in Table 6 in the Appendix. We derive three major findings from these results, which are similar to our previous findings for read disturb behavior in MLC NAND flash memory [16].

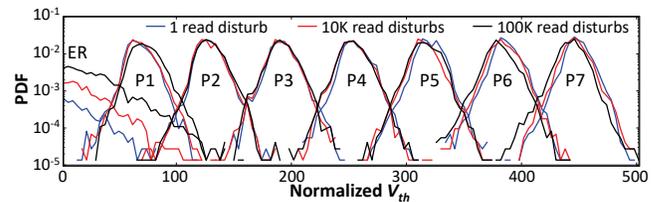

**Figure 19: Threshold voltage distribution for TLC NAND flash memory after 1, 10K, and 100K read disturb operations.**

First, as the read disturb count increases, the threshold voltages increase (i.e., the voltage distribution shifts to the right). In particular, we find that the distribution shifts are greater for lower-voltage states, indicating that read disturb impacts cells in the ER and P1 states the most. This is because we apply the same pass-through voltage ($V_{pass}$) to *all* unread cells during a read operation, *regardless* of the threshold voltages of the cells. A lower threshold voltage on a cell induces a larger voltage difference ($V_{pass} - V_{th}$) through the tunnel oxide layer of the cell, and in turn generates a stronger tunneling current, making the cell more vulnerable to read disturb (as described in detail in our prior work [16]).

Second, cells whose threshold voltages are closer to the point at which the voltage distributions of the ER and P1 states intersect are more vulnerable to read disturb errors. This is because process variation causes different cells to have different degrees of vulnerability to read disturb. We find that cells that are *prone* to read disturb end up at the right tail of the threshold voltage distribution of the ER state, as these cells' threshold voltages increase more rapidly, and that cells that are relatively *resistant* to read disturb end up at the left tail of the threshold voltage distribution of the P1 state, as their threshold voltages increase more slowly. We can exploit this divergent behavior of cells that end up at the left and right distribution tails to perform error recovery in the event of an uncorrectable error, as we discuss in Section 6.4.

Third, unlike with the other states, the threshold voltages of the cells at the left tail of the highest-voltage state (P7) in TLC NAND flash memory actually *decreases* as the read disturb count increases. This occurs for two reasons: (1) applying $V_{pass}$ causes electrons to move from the floating gate to the control gate for a cell at high voltage (i.e., a cell containing a large number of electrons), thus *reducing* its threshold voltage [16, 203]; and (2) some retention time elapses while we sweep the voltages during our read disturb experiments, inducing trap-assisted tunneling (see Section 4.4) and leading to retention errors that decrease the voltage.

More findings on the nature of read disturb and the impact of read disturb on NAND flash memory errors and lifetime can be found in our prior work [16].

### 4.6 Large-Scale Studies on SSD Errors

The error characterization studies we have discussed so far examine the susceptibility of real NAND flash memory devices to specific error sources, by conducting controlled experiments on individual flash devices in controlled environments. To examine the aggregate effect of these error sources on flash devices that operate in the field, several recent studies have analyzed the reliability of SSDs deployed at a large scale (i.e., tens to hundreds of thousands of SSDs) in production data centers [122, 142, 162]. Unlike the controlled low-level error characterization studies discussed in Sections 4.1 through 4.5, these large-scale studies analyze the observed errors and error rates in an *uncontrolled* manner, i.e., based on real data center workloads operating at field conditions as opposed to controlled access patterns and controlled conditions. As such, these large-scale studies can study flash memory behavior and reliability using only a black-box approach, where they are able to access only the registers used by the SSD to record select statistics. On the other hand, these studies incorporate the effects of a real system, including the system software stack and real workloads [122], on the flash memory devices, which is not present in the controlled small-scale studies.

These large-scale studies have made a number of observations across large sets of SSDs. We highlight five key observations from these studies. First, SSD failure rates do *not* increase monotonically with the P/E cycle count, and instead exhibit several distinct periods of reliability, where the failure rates between each period can vary by as much as 81.7% [122]. Second, the raw bit error rate grows with the age of the device even if the P/E cycle count is held constant, indicating that mechanisms such as silicon aging are likely contributing to the error rate [142]. Third, the observed failure rate of SSDs has been noted to be significantly higher than the failure rates specified by the manufacturers [162]. Fourth, higher operating temperatures can lead to higher failure rates, but modern SSDs employ throttling techniques that reduce the access rates to the underlying flash chips, which can greatly reduce the negative reliability impact of higher temperatures [122]. Fifth, while SSD failure rates are higher than specified, the overall occurrence of *uncorrectable* errors is lower than expected because (1) effective bad block management policies (see Section II-C) are implemented in SSD controllers; and (2) certain types of error sources, such as read disturb [122, 142] and incomplete erase operations [142], have yet to become a major source of uncorrectable errors at the system level.

## 5 ERROR MITIGATION

Several different types of errors can occur in NAND flash memory, as we described in Section 4. As NAND flash memory continues to scale to smaller technology nodes, the magnitude of these errors has been increasing [120, 153, 206]. This, in turn, uses up the limited error correction capability of ECC more rapidly than in past flash memory generations and shortens the lifetime of modern SSDs. To overcome the decrease in lifetime, a number of error mitigation techniques, which exploit intrinsic properties of the different types of errors to reduce the rate at which they lead to raw bit errors, have been designed. In this section, we discuss how the flash controller mitigates each of the error types via proposed error mitigation mechanisms. Table 3 shows the techniques we overview and which errors (from Section 4) they mitigate.

Table 3: List of different types of errors mitigated by NAND flash error mitigation mechanisms.

| Mitigation Mechanism | Error Type | | | | |
|---|---|---|---|---|---|
| | P/E Cycling [14, 15, 116] (§4.1) | Program [12, 116, 153] (§4.2) | Cell-to-Cell Interference [14, 18, 26, 108] (§4.3) | Data Retention [14, 17, 24, 25, 126] (§4.4) | Read Disturb [14, 16, 56, 126] (§4.5) |
| **Shadow Program Sequencing** [12, 18] (Section 5.1) | | | X | | |
| **Neighbor-Cell Assisted Error Correction** [26] (Section 5.2) | | | X | | |
| **Refresh** [24, 25, 128, 147] (Section 5.3) | | | | X | X |
| **Read-Retry** [15, 53, 201] (Section 5.4) | X | | | X | X |
| **Voltage Optimization** [16, 17, 75] (Section 5.5) | X | | | X | X |
| **Hot Data Management** [59, 60, 115] (Section 5.6) | X | X | X | X | X |
| **Adaptive Error Mitigation** [23, 36, 62, 190, 193] (Section 5.7) | X | X | X | X | X |

### 5.1 Shadow Program Sequencing

As discussed in Section 4.3, cell-to-cell program interference is a function of the distance between the cells of the wordline that is being programmed and the cells of the victim wordline. The impact of program interference is greatest on a victim wordline when either of the victim's immediately-adjacent wordlines is programmed (e.g., if we program WL1 in Figure 8, WL0 and WL2 experience the greatest amount of interference). Early MLC flash memories used one-shot programming, where both the LSB and MSB pages of a wordline are programmed at the same time. As flash memory scaled to smaller process technologies, one-shot programming resulted in much larger amounts of cell-to-cell program interference. As a result, manufacturers introduced two-step programming for MLC NAND flash (see Section 3.4), where the SSD controller writes values of the two pages within a wordline in two independent steps.

The SSD controller minimizes the interference that occurs during two-step programming by using *shadow program sequencing* [12, 18, 149] to determine the order that data is written to different pages in a block. If we program the LSB and MSB pages of the same wordline back to back, as shown in Figure 20a, both programming steps induce interference on a *fully-programmed wordline* (i.e., a wordline where both the LSB and MSB pages are already written). For example, if the controller programs both pages of WL1 back to back, shown as bold page programming operations in Figure 20a, the program operations induce a high amount of interference on WL0, which is fully programmed. The key idea of shadow program sequencing is to ensure that a fully-programmed wordline experiences interference minimally, i.e., *only* during MSB page programming (and *not* during LSB page programming). In shadow program sequencing, we assign a unique page number to each page within a block, as shown in Figure 20b. The LSB page of wordline *i* is numbered page $2i - 1$, and the MSB page is numbered page $2i + 2$. The only exceptions to the numbering are the LSB page of wordline 0 (page 0) and the MSB page of the last wordline *n* (page $2n + 1$). Two-step programming writes to pages in *increasing* order of page number inside a block [12, 18, 149], such that a *fully-programmed* wordline experiences interference only from the MSB page programming of the wordline directly above it, shown as the bold page programming operation in Figure 20b. With this programming order/sequence, the LSB page of the wordline above, and both pages of the wordline below, do *not* cause interference to fully-programmed data [12, 18, 149], as these two pages are programmed *before* programming the MSB page of the given wordline. Foggy-fine programming in TLC NAND flash (see Section 3.4) uses a similar ordering to reduce cell-to-cell program interference, as shown in Figure 20c.

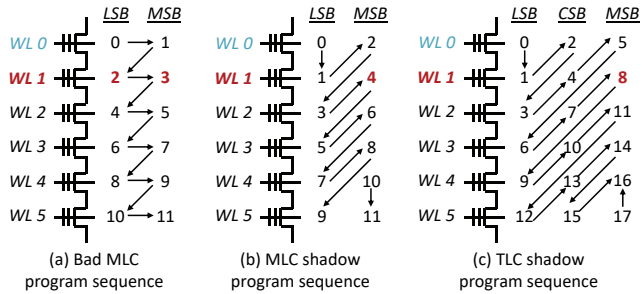

(a) Bad MLC program sequence  (b) MLC shadow program sequence  (c) TLC shadow program sequence

**Figure 20: Order in which the pages of each wordline (WL) are programmed using (a) a bad programming sequence, and using shadow sequencing for (b) MLC and (c) TLC NAND flash. The bold page programming operations for WL1 induce cell-to-cell program interference when WL0 is fully programmed.**

Shadow program sequencing is an effective solution to minimize cell-to-cell program interference on fully-programmed wordlines during two-step programming, and is employed in commercial SSDs today.

### 5.2 Neighbor-Cell Assisted Error Correction

The threshold voltage shift that occurs due to program interference is highly correlated with the values stored in the cells of the *immediately-adjacent wordlines*, as we discussed in Section 4.3. Due to this correlation, knowing the value programmed in the immediately-adjacent cell (i.e., a *neighbor cell*) makes it easier to correctly determine the value stored in the flash cell that is being read [26]. We describe a recently proposed error correction method that takes advantage of this observation, called *neighbor-cell-assisted error correction* (NAC). The key idea of NAC is to use the data values stored in the cells of the immediately-adjacent wordline to determine a better set of read reference voltages for the wordline that is being read. Doing so leads to a more accurate identification of the logical data value that is being read, as the data in the immediately-adjacent wordline was *partially responsible* for shifting the threshold voltage of the cells in the wordline that is being read when the immediately-adjacent wordline was programmed.

Figure 21 shows an operational example of NAC that is applied to eight bitlines (BL) of an MLC flash wordline. The SSD controller first reads a flash page from a wordline using the standard read reference voltages (step 1 in Figure 21). The bit values read from the wordline are then buffered in the controller. If there are no errors uncorrectable by ECC, the read was successful, and nothing else is done. However, if there are errors that are *uncorrectable* by ECC, we assume that the threshold voltage distribution of the page shifted due to cell-to-cell program interference, triggering further correction. In this case, NAC reads the LSB and MSB pages of the wordline *immediately above* the requested page (i.e., the *adjacent* wordline that was programmed *after* the requested page) to classify the cells of the requested page (step 2). NAC then identifies the cells adjacent to (i.e., connected to the same bitline as) the ER cells (i.e., cells in the immediately above wordline that are in the ER state), such as the cells on BL1, BL3, and BL7 in Figure 21. NAC rereads these cells using read reference voltages that *compensate for* the threshold voltage shift caused by programming the adjacent cell to the ER state (step 3). If ECC can correct the remaining errors, the controller returns the corrected page to the host. If ECC fails again, the process is repeated using a different set of read reference voltages for cells that are adjacent to the P1 cells (step 4). If ECC continues to fail, the process is repeated for cells that are adjacent to P2 and P3 cells (steps 5 and 6, respectively, which are not shown in the figure) until either ECC is able to correct the page or all possible adjacent values are exhausted.

|  |  | BL0 | BL1 | BL2 | BL3 | BL4 | BL5 | BL6 | BL7 |
|---|---|---|---|---|---|---|---|---|---|
|  | *Originally-programmed value* | 11 | 00 | 01 | 10 | 11 | 00 | 01 | 00 |
|  | 1. Read (using $V_{opt}$) with *errors* | 01 | 00 | 00 | 00 | 11 | 10 | 00 | 01 |
| N | 2. Read **adjacent wordline** | P2 | ER | P2 | ER | P1 | P3 | P1 | ER |
| A | 3. *Correct cells* adjacent to ER | 01 | 00 | 00 | **10** | 11 | 10 | 00 | **00** |
| C | 4. *Correct cells* adjacent to P1 | 01 | 00 | 00 | 10 | 11 | 10 | **01** | 00 |

**Figure 21: Overview of neighbor-cell-assisted error correction (NAC).**

NAC extends the lifetime of an SSD by reducing the number of errors that need to be corrected using the limited correction capability of ECC. With the use of experimental data collected from real MLC NAND flash memory chips, we show that NAC extends the NAND flash memory lifetime by 33% [26]. Our previous work [26] provides a detailed description of NAC, including a theoretical treatment of why it works and a practical implementation that minimizes the number of reads performed, even in the case when the neighboring wordline itself has errors.

## 5.3 Refresh Mechanisms

As we see in Figure 12, during the time period after a flash page is programmed, retention (Section 4.4) and read disturb (Section 4.5) can cause an increasing number of raw bit errors to accumulate over time. This is particularly problematic for a page that is not updated frequently. Due to the limited error correction capability, the accumulation of these errors can potentially lead to data loss for a page with a *high retention age* (i.e., a page that has not been programmed for a long time). To avoid data loss, *refresh mechanisms* have been proposed, where the stored data is periodically read, corrected, and reprogrammed, in order to eliminate the retention and read disturb errors that have accumulated prior to this periodic read/correction/reprogramming (i.e., refresh). The concept of refresh in flash memory is thus conceptually similar to the refresh mechanisms found in DRAM [30, 73, 112, 113]. By performing refresh and limiting the number of retention and read disturb errors that can accumulate, the lifetime of the SSD increases significantly. In this section, we describe three types of refresh mechanisms used in modern SSDs: remapping-based refresh, in-place refresh, and read reclaim.

**Remapping-Based Refresh.** Flash cells must first be erased before they can be reprogrammed, due to the fact the programming a cell via ISPP can only increase the charge level of the cell but not reduce it (Section 3.4). The key idea of *remapping-based refresh* is to periodically read data from each valid flash block, correct any data errors, and *remap the data to a different physical location*, in order to prevent the data from accumulating too many retention errors [10, 24, 25, 128, 147]. During each refresh interval, a block with valid data that needs to be refreshed is selected. The valid data in the selected block is read out page by page and moved to the SSD controller. The ECC engine in the SSD controller corrects the errors in the read data, including retention errors that have accumulated since the last refresh. A new block is then selected from the free list (see Section 2.3), the error-free data is programmed to a page within the new block, and the logical address is remapped to point to the newly-programmed physical page. By reducing the accumulation of retention and read disturb errors, remapping-based refresh increases SSD lifetime by an average of 9x for a variety of disk workloads [24, 25].

Prior work proposes extensions to the basic remapping-based refresh approach. One work, *refresh SSDs*, proposes a refresh scheduling algorithm based on an earliest deadline first policy to guarantee that all data is refreshed in time [128]. The *quasi-nonvolatile SSD* proposes to use remapping-based refresh to choose between improving flash endurance and reducing the flash programming latency (by using larger ISPP step-pulses) [147]. In the quasi-nonvolatile SSD, refresh requests are deprioritized, scheduled at idle times, and can be interrupted after refreshing any page within a block, to minimize the delays that refresh can cause for the response time of pending workload requests to the SSD. A refresh operation can also be triggered proactively based on the data read latency observed for a page, which is indicative of how many errors the page has experienced [19]. Triggering refresh *proactively* based on the observed read latency (as opposed to doing so *periodically*) improves SSD latency and throughput [19]. Whenever the read latency for a page within a block exceeds a fixed threshold, the valid data in the block is refreshed, i.e., remapped to a new block [19].

**In-Place Refresh.** A major drawback of remapping-based refresh is that it performs *additional writes* to the NAND flash memory, accelerating wearout. To reduce the wearout overhead of refresh, we propose *in-place refresh* [10, 24, 25]. As data sits unmodified in the SSD, data retention errors dominate [14, 25, 177], leading to charge loss and causing the threshold voltage distribution to shift to the left, as we showed in Section 4.4. The key idea of in-place refresh is to incrementally replenish the lost charge of each page *at its current location*, i.e., in place, without the need for remapping.

Figure 22 shows a high-level overview of in-place refresh for a wordline. The SSD controller first reads all of the pages in the wordline (❶ in Figure 22). The controller invokes the ECC decoder to correct the errors within each page (❷), and sends the corrected data back to the flash chips (❸). In-place refresh then invokes a modified version of the ISPP mechanism (see Section 3.4), which we call *Verify-ISPP* (V-ISPP), to compensate for retention errors by restoring the charge that was lost. In V-ISPP, we first verify the voltage currently programmed in a flash cell (❹). If the current voltage of the cell is *lower* than the target threshold voltage of the state that the cell should be in, V-ISPP pulses the programming voltage in steps, gradually injecting charge into the cell until the cell returns to the target threshold voltage (❺). If the current voltage of the cell is *higher* than the target threshold voltage, V-ISPP inhibits the programming pulses to the cell.

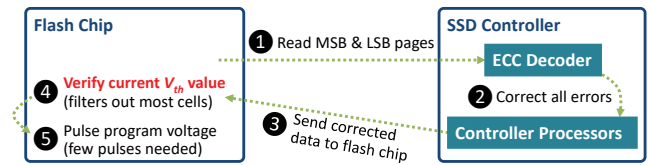

Figure 22: Overview of in-place refresh mechanism for MLC NAND flash memory.

When the controller invokes in-place refresh, it is unable to use shadow program sequencing (Section 5.1), as all of the pages within the wordline have already been programmed. However, unlike traditional ISPP, V-ISPP does not introduce a high amount of cell-to-cell program interference (Section 4.3) for two reasons. First, V-ISPP programs *only* those cells that have retention errors, which typically account for less than 1% of the total number of cells in a wordline selected for refresh [24]. Second, for the small number of cells that are selected to be refreshed, their threshold voltage is usually only slightly lower than the target threshold voltage, which means that only a few programming pulses need to be applied. As cell-to-cell interference is linearly correlated with the threshold voltage change to immediately-adjacent cells [18, 26], the small voltage change on these in-place refreshed cells leads to only a small interference effect.

One issue with in-place refresh is that it is unable to correct retention errors for cells in lower-voltage states. Retention errors cause the threshold voltage of a cell in a lower-voltage state to *increase* (e.g., see Section 4.4, ER and P1 states in Figure 18), but V-ISPP *cannot* decrease the threshold voltage of a cell. To achieve a balance between the wearout overhead due to remapping-based refresh and errors that increase the threshold voltage due to in-place refresh, we propose *hybrid in-place refresh* [10, 24, 25]. The key idea is to use in-place refresh when the number of program errors (caused due to reprogramming) is within the correction capability of ECC, but to use remapping-based refresh if the number of program errors is too large to tolerate. To accomplish this, the controller tracks the number of *right-shift errors* (i.e., errors that move a cell to a higher-voltage state) [24, 25]. If the number of right-shift errors remains under a certain threshold, the controller performs in-place refresh; otherwise, it performs remapping-based refresh. Such a hybrid in-place refresh mechanism increases SSD lifetime by an average of 31x for a variety of disk workloads [24, 25].

**Read Reclaim to Reduce Read Disturb Errors.** We can also mitigate read disturb errors using an idea similar to remapping-based refresh, known as *read reclaim*. The key idea of read reclaim is to remap the data in a block to a new flash block, if the block has experienced a high number of reads [59, 60, 88]. To bound the number of read disturb errors, some flash vendors specify a maximum number of tolerable reads for a flash block, at which point read reclaim rewrites the data to a new block (just as is done for remapping- based refresh).

**Adaptive Refresh and Read Reclaim Mechanisms.** For the refresh and read reclaim mechanisms discussed above, the SSD controller can (1) invoke the mechanisms at fixed regular intervals; or (2) *adapt* the rate at which it invokes the mechanisms, based on various conditions that impact the rate at which data retention and read disturb errors occur. By adapting the mechanisms based on the current conditions of the SSD, the controller can reduce the overhead of performing refresh or read reclaim. The controller can adaptively adjust the rate that the mechanisms are invoked based on (1) the wearout (i.e., the current P/E cycle count) of the NAND flash memory [24, 25]; or (2) the temperature of the SSD [14, 17].

As we discuss in Section 4.4, for data with a given retention age, the number of retention errors grows as the P/E cycle count increases. Exploiting this P/E cycle dependent behavior of retention time, the SSD controller can perform refresh less frequently (e.g., once every year) when the P/E cycle count is low, and more frequently (e.g., once every week) when the P/E cycle count is high, as proposed and described in our prior works [24, 25]. Similarly, for data with a given read disturb count, as the P/E cycle count increases, the number of read disturb errors increases as well [16]. As a result, the SSD controller can perform read reclaim less frequently (i.e., it increases the maximum number of tolerable reads per block before read reclaim is triggered) when the P/E cycle count is low, and more frequently when the P/E cycle count is high.

Prior works demonstrate that for a given retention time, the number of data retention errors increases as the NAND flash memory's operating temperature increases [14, 17]. To compensate for the increased number of retention errors at high temperature, a state-of-the-art SSD controller adapts the rate at which it triggers refresh. The SSD contains sensors that monitor the current environmental temperature every few milliseconds [122, 187]. The controller then uses the Arrhenius equation [3, 128, 199] to estimate the rate at which retention errors accumulate at the current temperature of the SSD. Based on the error rate estimate, the controller decides if it needs to increase the rate at which it triggers refresh to ensure that the data is not lost.

By employing adaptive refresh and/or read reclaim mechanisms, the SSD controller can successfully reduce the mechanism overheads while effectively mitigating the larger number of data retention errors that occur under various conditions.

### 5.4 Read-Retry

In earlier generations of NAND flash memory, the read reference voltage values were fixed at design time [15, 126]. However, several types of errors cause the threshold voltage distribution to shift, as shown in Figure 13. To compensate for threshold voltage distribution shifts, a mechanism called *read-retry* has been implemented in modern flash memories (typically those below 30 nm for planar flash [15, 53, 166, 201]).

The read-retry mechanism allows the read reference voltages to dynamically adjust to changes in distributions. During read-retry, the SSD controller first reads the data out of NAND flash memory with the default read reference voltage. It then sends the data for error correction. If ECC successfully corrects the errors in the data, the read operation succeeds. Otherwise, the SSD controller reads the memory again with a *different* read reference voltage. The controller repeats these steps until it either successfully reads the data using a certain set of read reference voltages or is unable to correctly read the data using all of the read reference voltages that are available to the mechanism.

While read-retry is widely implemented today, it can significantly increase the overall read operation latency due to the multiple read attempts it causes [17]. Mechanisms have been proposed to reduce the number of read-retry attempts while taking advantage of the effective capability of read-retry for reducing read errors, and read-retry has also been used to enable mitigation mechanisms for various other types of errors, as we describe in Section 5.5. As a result, read-retry is an essential mechanism in modern SSDs to mitigate read errors (i.e., errors that manifest themselves during a read operation).

### 5.5 Voltage Optimization

Many raw bit errors in NAND flash memory are affected by the various voltages used within the memory to enable reading of values. We give two examples. First, a suboptimal *read reference voltage* can lead to a large number of read errors (Section 4), especially after the threshold voltage distribution shifts. Second, as we saw in Section 4.5, the *pass-through voltage* can have a significant effect on the number of read disturb errors that occur. As a result, optimizing these voltages such that they minimize the total number of errors that are induced can greatly mitigate error counts. In this section, we discuss mechanisms that can discover and employ the optimal[2] read reference and pass-through voltages.

**Optimizing Read Reference Voltages Using Disparity-Based Approximation and Sampling.** As we discussed in Section 5.4, when the threshold voltage distribution shifts, it is important to move the read reference voltage to the point where the number of read errors is minimized. After the shift occurs and the threshold voltage distribution of each state widens, the distributions of different states may overlap with each other, causing many of the cells within the overlapping regions to be misread. The number of errors due to misread cells can be minimized by setting the read reference voltage to be exactly at the point where the distributions of two neighboring states intersect, which we call the *optimal read reference voltage* ($V_{opt}$) [17, 18, 26, 116, 148], illustrated in Figure 23. Once the optimal read reference voltage is applied, the raw bit error rate is minimized, improving the reliability of the device.

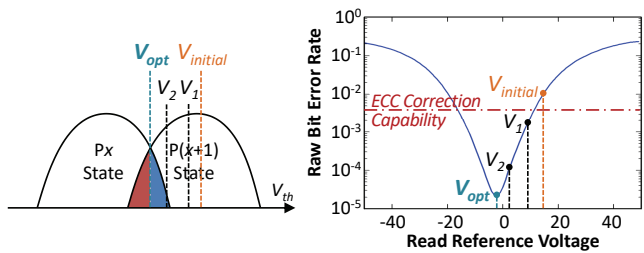

**Figure 23: Finding the optimal read reference voltage after the threshold voltage distributions overlap (left), and raw bit error rate as a function of the selected read reference voltage (right).**

One approach to finding $V_{opt}$ is to adaptively learn and apply the optimal read reference voltage for each flash block through sampling [17, 37, 42, 197]. The key idea is to periodically (1) use *disparity* information (i.e., the ratio of 1s to 0s in the data) to attempt to find a read reference voltage for which the error rate is lower than the ECC correction capability; and to (2) use *sampling* to efficiently tune the read reference voltage to its optimal value to reduce the read operation latency. Prior characterization of real NAND flash memory [17, 148] found that the value of $V_{opt}$ does *not* shift greatly over a short period of time (e.g., a day), and that all pages within a block experience *similar* amounts of threshold voltage shifts, as they have the same amount of wearout and are programmed around the same time [17, 148]. Therefore, we can invoke our $V_{opt}$ learning mechanism periodically (e.g., daily) to efficiently tune the *initial read reference voltage* (i.e., the first read reference voltage used when the controller invokes the read-retry mechanism, described in Section 5.4) for each flash block, ensuring that the initial voltage used by read-retry stays close to $V_{opt}$ even as the threshold voltage distribution shifts.

The SSD controller searches for $V_{opt}$ by counting the number of errors that need to be corrected by ECC during a read. However, there may be times where the initial read reference voltage ($V_{initial}$) is set to a value at which the number of errors during a read exceeds the ECC correction capability, such as the raw bit error rate for $V_{initial}$ in Figure 23 (right). When the ECC correction capability is exceeded, the SSD controller is unable to count how many errors exist in the raw data. The SSD controller uses *disparity-based read reference voltage approximation* [37, 42, 197] for each flash block to try to bring $V_{initial}$ to a region where the number of errors does not exceed the ECC correction capability. Disparity-based read reference voltage approximation takes advantage of data scrambling. Recall from Section 2.3 that to minimize data value dependencies for the error rate, the SSD controller scrambles the data written to the SSD to probabilistically ensure that an equal number of 0s and 1s exist in the flash memory cells. The key idea of disparity-based read reference voltage approximation is to find the read reference voltages that result in approximately 50% of the cells reading out bit value 0, and the other 50% of the cells reading out bit value 1. To achieve this, the SSD controller employs a binary search algorithm, which tracks the ratio of 0s to 1s for each read reference voltage it tries. The binary search tests various read reference voltage values, using the ratios of previously tested voltages to narrow down the range where the read reference voltage can have an equal ratio of 0s to 1s. The binary search algorithm continues narrowing down the range until it finds a read reference voltage that satisfies the ratio.

---

[2]Or, more precisely, near-optimal, if the read-retry steps are too coarse grained to find the optimal voltage.

The usage of the binary search algorithm depends on the type of NAND flash memory used within the SSD. For SLC NAND flash, the controller searches for only a single read reference voltage. For MLC NAND flash, there are three read reference voltages: the LSB is determined using $V_b$, and the MSB is determined using both $V_a$ and $V_c$ (see Section 3.3). Figure 24 illustrates the search procedure for MLC NAND flash. First, the controller uses binary search to find $V_b$, choosing a voltage that reads the LSB of 50% of the cells as data value 0 (step 1 in Figure 24). For the MSB, the controller uses the discovered $V_b$ value to help search for $V_a$ and $V_c$. Due to scrambling, cells should be equally distributed across each of the four voltage states. The controller uses binary search to set $V_a$ such that 25% of the cells are in the ER state, by ensuring that half of the cells *to the left of* $V_b$ are read with an MSB of 0 (step 2). Likewise, the controller uses binary search to set $V_c$ such that 25% of the cells are in the P3 state, by ensuring that half of the cells *to the right of* $V_b$ are read with an MSB of 0 (step 3). This procedure is extended in a similar way to approximate the voltages for TLC NAND flash.

If disparity-based approximation finds a value for $V_{initial}$ where the number of errors during a read can be counted by the SSD controller, the controller invokes *sampling-based adaptive $V_{opt}$ discovery* [17] to minimize the error count, and thus reduce the read latency. Sampling-based adaptive $V_{opt}$ discovery learns and records $V_{opt}$ for the *last-programmed page* in each block. We sample only the last-programmed page because it is the page with the lowest data retention age in the flash block. As retention errors cause the higher-voltage states to shift to the left (i.e., to lower voltages), the last-programmed page usually provides an *upper bound* of $V_{opt}$ for the entire block.

During sampling-based adaptive $V_{opt}$ discovery, the SSD controller first reads the last-programmed page using $V_{initial}$, and attempts to correct the errors in the raw data read from the page.

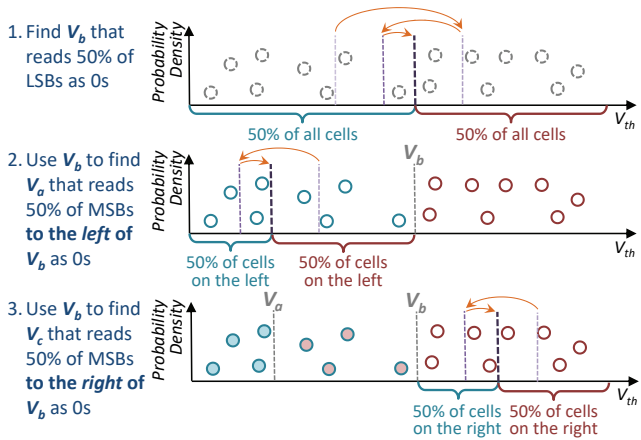

**Figure 24: Disparity-based read reference voltage approximation to find $V_{initial}$ for MLC NAND flash memory. Each circle represents a cell, where a dashed border indicates that the LSB is undetermined, a solid border indicates that the LSB is known, a hollow circle indicates that the MSB is unknown, and a filled circle indicates that the MSB is known.**

Next, it records the number of raw bit errors as the current lowest error count $N_{ERR}$, and sets the applied read reference voltage ($V_{ref}$) as $V_{initial}$. Since $V_{opt}$ typically decreases over retention age, the controller first attempts to lower the read reference voltage for the last-programmed page, decreasing the voltage to $V_{ref} - \Delta V$ and reading the page. If the number of corrected errors in the new read is less than or equal to the old $N_{ERR}$, the controller updates $N_{ERR}$ and $V_{ref}$ with the new values. The controller continues to lower the read reference voltage until the number of corrected errors in the data is greater than the old $N_{ERR}$ or the lowest possible read reference voltage is reached. Since the optimal threshold voltage might increase in rare cases, the controller also tests increasing the read reference voltage. It increases the voltage to $V_{ref} + \Delta V$ and reads the last-programmed page to see if $N_{ERR}$ decreases. Again, it repeats increasing $V_{ref}$ until the number of corrected errors in the data is greater than the old $N_{ERR}$ or the highest possible read reference voltage is reached. The controller sets the initial read reference voltage of the block as the value of $V_{ref}$ at the end of this process so that the next time an uncorrectable error occurs, read-retry starts at a $V_{initial}$ that is hopefully closer to the optimal read reference voltage ($V_{opt}$).

During the course of the day, as more retention errors (the dominant source of errors on already-programmed blocks) accumulate, the threshold voltage distribution shifts to the left (i.e., voltages decrease), and our initial read reference voltage (i.e., $V_{initial}$) is now an upper bound for the read-retry voltages. Therefore, whenever read-retry is invoked, the controller now needs to only decrease the read reference voltages (as opposed to traditional read-retry, which tries *both* lower and higher voltages [17]). Sampling-based adaptive $V_{opt}$ discovery improves the *endurance* (i.e., the number of P/E cycles before the ECC correction capability is exceeded) of the NAND flash memory by 64% and reduces error correction latency by 10% [17], and is employed in some modern SSDs today.

**Other Approaches to Optimizing Read Reference Voltages.** One drawback of the sampling-based adaptive technique is that it requires time and storage overhead to find and record the per-block initial voltages. To avoid this, the SSD controller can employ an accurate *online threshold voltage distribution model* [10, 15, 116], which can efficiently track and predict the shift in the distribution over time. The model represents the threshold voltage distribution of each state as a probability density function (PDF), and the controller can use the model to calculate the intersection of the different PDFs. The controller uses the PDF in place of the threshold voltage sampling, determining $V_{opt}$ by calculating the intersection of the distribution of each state in the model. The endurance improvement from our state-of-the-art model-based $V_{opt}$ estimation technique [116] is within 2% of the improvement from an ideal $V_{opt}$ identification mechanism [116]. An online threshold voltage distribution model can be used for a number of other purposes, such as estimating the future growth in the raw bit error rate and improving error correction [116].

Other prior work examines adapting read reference voltages based on P/E cycle count, retention age, or read disturb. In one such work, the controller periodically learns read reference voltages by testing three read reference voltages on six pages per block, which the work demonstrates to be sufficiently accurate [148]. Similarly, error correction using LDPC soft decoding (see Section 6.2) requires reading the same page using multiple sets of read reference voltages to provide fine-grained information on the probability of each cell representing a bit value 0 or a bit value 1. Another prior work optimizes the read reference voltages to increase the ECC correction capability without increasing the coding rate [184].

**Optimizing Pass-Through Voltage to Reduce Read Disturb Errors.** As we discussed in Section 4.5, the vulnerability of a cell to read disturb is directly correlated with the voltage difference ($V_{pass} - V_{th}$) through the cell oxide [16]. Traditionally, a single $V_{pass}$ value is used *globally* for the entire flash memory, and the value of $V_{pass}$ must be higher than *all* potential threshold voltages within the chip to ensure that unread cells along a bitline are turned on during a read operation (see Section 3.3). To reduce the impact of read disturb, we can tune $V_{pass}$ to reduce the size of the voltage difference ($V_{pass} - V_{th}$). However, it is difficult to reduce $V_{pass}$ globally, as any cell with a value of $V_{th} > V_{pass}$ introduces an error during a read operation (which we call a *pass-through error*).

We propose a mechanism that can dynamically lower $V_{pass}$ while ensuring that it can correct any new pass-through errors introduced. The key idea of the mechanism is to lower $V_{pass}$ only for those blocks where ECC has enough leftover error correction capability (see Section 2.3) to correct the newly introduced pass-through errors. When the retention age of the data within a block is low, we find that the raw bit error rate of the block is much lower than the rate for the block when the retention age is high, as the number of data retention and read disturb errors remains low at low retention age [16, 60]. As a result, a block with a low retention age has significant *unused* ECC correction capability, which we can use to correct the pass-through errors we introduce when we lower $V_{pass}$, as shown in Figure 25. Thus, when a block has a low retention age, the controller lowers $V_{pass}$ aggressively, making it much less likely for read disturbs to induce an uncorrectable error.

When a block has a high retention age, the controller also lowers $V_{pass}$, but does not reduce the voltage aggressively, since the limited ECC correction capability now needs to correct retention errors, and might not have enough unused correction capability to correct many new pass-through errors. By reducing $V_{pass}$ aggressively when a block has a low retention age, we can extend the time before the ECC correction capability is exhausted, improving the flash lifetime.

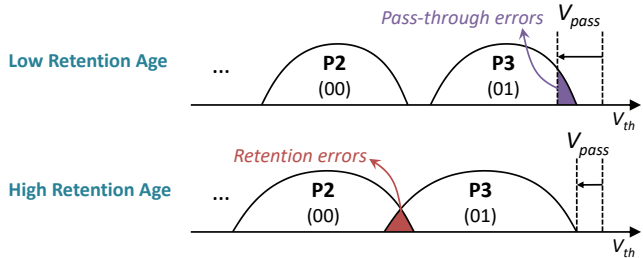

Figure 25: Dynamic pass-through voltage tuning at different retention ages.

Our read disturb mitigation mechanism [16] learns the minimum pass-through voltage for each block, such that all data within the block can be read correctly with ECC. Our learning mechanism works online and is triggered periodically (e.g., daily). The mechanism is implemented in the controller, and has two components. It first finds the size of the ECC margin $M$ (i.e., the unused correction capability) that can be exploited to tolerate additional read errors for each block. Once it knows the available margin $M$, our mechanism calibrates $V_{pass}$ on a per-block basis to find the lowest value of $V_{pass}$ that introduces no more than $M$ additional raw errors (i.e., there are no more than M cells where $V_{th} > V_{pass}$). Our findings on MLC NAND flash memory show that the mechanism can improve flash endurance by an average of 21% for a variety of disk workloads [16].

**Programming and Erase Voltages.** Prior work also examines tuning the programming and erase voltages to extend flash endurance [75]. By decreasing the two voltages when the P/E cycle count is low, the accumulated wearout for each program or erase operation is reduced, which, in turn, increases the overall flash endurance. Decreasing the programming voltage, however, comes at the cost of increasing the time required to perform ISPP, which, in turn, increases the overall SSD write latency [75].

### 5.6 Hot Data Management

The data stored in an SSD can be accessed by the host at different rates. For example, we find that across a wide range of disk workloads, almost 100% of the write operations target less than 1% of the pages within an SSD [115], exhibiting high temporal write locality. We call the frequently-written subset of pages *write-hot* pages. Likewise, pages with a high amount of temporal read locality are called *read-hot* pages. A number of issues can arise when an SSD does not distinguish between write-hot pages and *write-cold* pages (i.e., pages with low temporal write locality), or between read-hot pages and *read-cold* pages (i.e., pages with low temporal read locality). For example, if write-hot pages and write-cold pages are kept within the same block, intelligent refresh mechanisms cannot avoid refreshes to pages that were overwritten recently, increasing not only energy consumption but also write amplification due to remapping-based refresh [115]. Likewise, if read-hot and read-cold pages are kept within the same block, read-cold pages are unnecessarily exposed to a high number of read disturb errors [59, 60]. *Hot data management* refers to a set of mechanisms that can identify write-hot or read-hot pages in the SSD. The key idea is to apply special SSD management policies by placing hot pages and cold pages into *separate* flash blocks.

*Write-hotness aware refresh management* (WARM) [115] efficiently identifies write-hot pages, and designates a small pool of blocks in the SSD to exclusively store write-hot data. As write-hot data is overwritten more frequently than the refresh interval, the SSD controller can skip refresh operations to the write-hot blocks. WARM reduces the write amplification overhead of refresh, which translates to an average lifetime improvement of 21% over a state-of-the-art refresh mechanism across a range of disk workloads [115]. Another work [183] proposes to reuse the correctly functioning flash pages within bad blocks (see Section 2.3) to store write-cold data. This technique increases the total number of usable blocks available for overprovisioning, and extends flash lifetime by delaying the point at which each flash chip reaches the upper limit of bad blocks it can tolerate.

RedFTL identifies and replicates read-hot pages across multiple flash blocks, allowing the controller to evenly distribute read requests to these pages across the replicas [59]. Other works reduce the number of read reclaims (see Section 5.3) that need to be performed by mapping read-hot data to particular flash blocks and lowering the maximum possible threshold voltage for such blocks [21, 60]. By lowering the maximum possible threshold voltage for these blocks, the SSD controller can use a lower $V_{pass}$ value (see Section 5.5) on the blocks without introducing any additional errors during a read operation. To lower the maximum threshold voltage in these blocks, the width of the voltage window for each voltage state is decreased, and each voltage window shifts to the left [21, 60]. Another work applies stronger ECC encodings to *only* read-hot blocks based on the total read count of the block, in order to increase SSD endurance without significantly reducing the amount of overprovisioning [20] (see Section 2.4 for a discussion on the tradeoff between ECC strength and overprovisioning).

### 5.7 Adaptive Error Mitigation Mechanisms

Due to the many different factors that contribute to raw bit errors, error rates in NAND flash memory can be highly variable. Adaptive error mitigation mechanisms are capable of adapting error tolerance capability to the error rate. They provide stronger error tolerance capability when the error rate is higher, improving flash lifetime significantly. When the error rate is low, adaptive error mitigation techniques reduce error tolerance capability to lower the cost of the error mitigation techniques. In this section, we examine two types of adaptive techniques: (1) multi-rate ECC and (2) dynamic cell levels.

**Multi-Rate ECC.** Some works propose to employ multiple ECC algorithms in the SSD controller [23, 36, 62, 69, 193]. Recall from Section 2.4 that there is a tradeoff between ECC strength (i.e., the coding rate; see Section 2.3) and overprovisioning, as a codeword (which contains a data chunk *and* its corresponding ECC information) uses more bits when stronger ECC is employed. The key idea of multi-rate ECC is to employ a weaker codeword (i.e., one that uses fewer bits for ECC) when the SSD is relatively new and has a smaller number of raw bit errors, and to use the saved SSD space to provide additional overprovisioning, as shown in Figure 26.

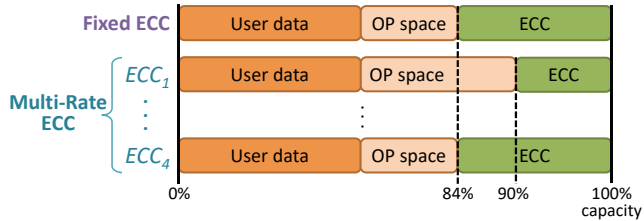

Figure 26: Comparison of space used for user data, overprovisioning, and ECC between a fixed ECC and a multi-rate ECC mechanism.

Let us assume that the controller contains a configurable ECC engine that can support $n$ different types of ECC codewords, which we call $ECC_i$. Figure 26 shows an example of multi-rate ECC that uses four ECC engines, where $ECC_1$ provides the weakest protection but has the smallest codeword, while $ECC_4$ provides the strongest protection with the largest codeword. We need to ensure that the NAND flash memory has enough space to fit the largest codewords, e.g., those for $ECC_4$ in Figure 26. Initially, when the raw bit error rate (RBER) is low, the controller employs $ECC_1$, as shown in Figure 27. The smaller codeword size for $ECC_1$ provides additional space for overprovisioning, as shown in Figure 26, and thus reduces the effects of write amplification. Multi-rate ECC works on an interval-by-interval basis. Every interval (in this case, a predefined number of P/E cycles), the controller measures the RBER. When the RBER exceeds the threshold set for transitioning from a weaker ECC to a stronger ECC, the controller switches to the stronger ECC. For example, when the SSD exceeds the first RBER threshold for switching ($T_1$ in Figure 27), the controller starts switching from $ECC_1$ to $ECC_2$. When switching between ECC engines, the controller uses the $ECC_1$ engine to decode data the next time the data is read out, and stores a new codeword using the $ECC_2$ engine. This process is repeated during the lifetime of flash memory for each stronger engine $ECC_i$, where each engine has a corresponding threshold that triggers switching [23, 36, 62], as shown in Figure 27.

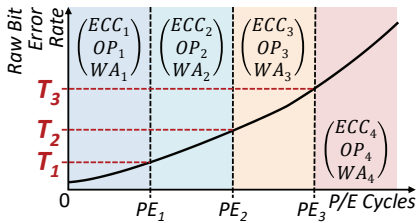

Figure 27: Illustration of how multi-rate ECC switches to different ECC codewords (i.e., $ECC_i$) as the RBER grows. $OP_i$ is the overprovisioning factor used for engine $ECC_i$, and $WA_i$ is the resulting write amplification value.

Multi-rate ECC allows the same maximum P/E cycle count for each block as if $ECC_n$ was used throughout the lifetime of the SSD, but reduces write amplification and improves performance during the periods where the lower strength engines are employed, by providing additional overprovisioning (see Section 2.4) during those times. As the lower-strength engines use smaller codewords (e.g., $ECC_1$ versus $ECC_4$ in Figure 26), the resulting free space can instead be employed to further increase the amount of overprovisioning within the NAND flash memory, which in turn increases the total lifetime of the SSD. We compute the lifetime improvement by modifying Equation 4 (Section 2.4) to account for each engine, as follows:

$$\text{Lifetime} = \sum_{i=1}^{n} \frac{\text{PEC}_i \times (1 + \text{OP}_i)}{365 \times \text{DWPD} \times \text{WA}_i \times R_{compress}} \quad (9)$$

In Equation 9, $WA_i$ and $OP_i$ are the write amplification and overprovisioning factor for $ECC_i$, and $PEC_i$ is the number of P/E cycles that $ECC_i$ is used for. Manufacturers can set parameters to maximize SSD lifetime in Equation 9, by optimizing the values of $WA_i$ and $OP_i$.

Figure 28 shows the lifetime improvements for a four-engine multi-rate ECC, with the coding rates for the four ECC engines ($ECC_1$–$ECC_4$) set to 0.90, 0.88, 0.86, and 0.84 (recall that a *lower* coding rate provides stronger protection; see Section 2.4), over a fixed ECC engine that employs a coding rate of 0.84. We see that the lifetime improvements of using multi-rate ECC are: (1) significant, with a 31.2% increase if the baseline NAND flash memory has 15% overprovisioning; and (2) greater when the SSD initially has a smaller amount of overprovisioning.

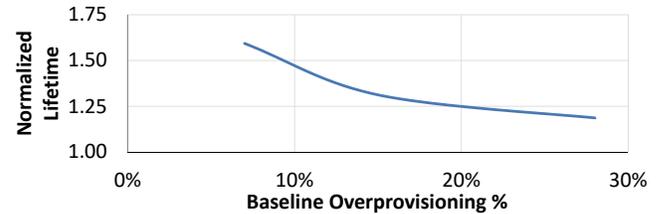

Figure 28: Lifetime improvements of using multi-rate ECC over using a fixed ECC coding rate.

**Dynamic Cell Levels.** A major reason that errors occur in NAND flash memory is because the threshold voltage distribution of each state overlaps more with those of neighboring states as the distributions widen over time. Distribution overlaps are a greater problem when more states are encoded within the same voltage range. Hence, TLC flash has a much lower endurance than MLC, and MLC has a much lower endurance than SLC (assuming the same process technology node). If we can increase the margins between the states' threshold voltage distributions, the amount of overlap can be reduced significantly, which in turn reduces the number of errors.

Prior work proposes to increase margins by *dynamically* reducing the number of bits stored within a cell, e.g., by going from three bits that encode eight states (TLC) to two bits that encode four states (equivalent to MLC), or to one bit that encodes two states (equivalent to SLC) [21, 190]. Recall that TLC uses the ER state and states P1–P7, which are spaced out approximately equally. When we *downgrade* a flash block (i.e., reduce the number of states its cells can represent) from eight states to four, the cells in the block now employ only the ER state and states P3, P5, and P7. As we can see from Figure 29, this provides large margins between states P3, P5, and P7, and provides an even larger margin between ER and P3. The SSD controller maintains a list of all of the blocks that have been downgraded. For each read operation, the SSD controller checks if the target block is in the downgraded block list, and uses this information to interpret the data that it reads out from the wordline of the block.

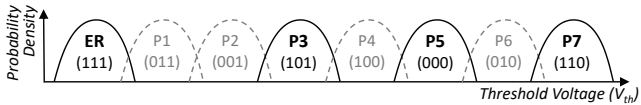

**Figure 29: States used when a TLC cell (with 8 states) is downgraded to an MLC cell (with 4 states).**

A cell can be downgraded to reduce various types of errors (e.g., wearout, read disturb). To reduce wearout, a cell is downgraded when it has high wearout. To reduce read disturb, a cell can be downgraded if it stores *read-hot* data (i.e., the most frequently read data in the SSD). By using fewer states for a block that holds read-hot data, we can reduce the impact of read disturb because it becomes harder for the read disturb mechanism to affect the distributions enough for them to overlap. As an optimization, the SSD controller can employ various hot-cold data partitioning mechanisms (e.g., [20, 21, 59, 115]) to keep read-hot data in specially designated blocks [20, 21, 59, 60], allowing the controller to reduce the size of the downgraded block list and isolate the impact of read disturb from *read-cold* (i.e., infrequently read) data.

Another approach to dynamically increasing the distribution margins is to perform program and erase operations more slowly when the SSD write request throughput is low [21, 75]. Slower program/erase operations allow the final voltage of a cell to be programmed more precisely, and reduce the amount of oxide degradation that occurs during programming. As a result, the distribution of each state is initially much narrower, and subsequent widening of the distributions results in much lower overlap for a given P/E cycle count. This technique improves the SSD lifetime by an average of 61.2% for a variety of disk workloads [75]. Unfortunately, the slower program/erase operations come at the cost of higher SSD latency, and are thus not applied during periods of high write traffic. One way to mitigate the impact of the higher write latency is to perform slower program/erase operations only during garbage collection, which ensures that the higher latency occurs only when the SSD is idle [21]. As a result, read and write requests from the host do not experience any additional delays.

## 6 ERROR CORRECTION AND DATA RECOVERY TECHNIQUES

Now that we have described a variety of error mitigation mechanisms that can target various types of error sources, we turn our attention to the error correction flow that is employed in modern SSDs as well as *data recovery techniques* that can be employed when the error correction flow fails to produce correct data.

Modern SSDs typically employ one of two types of ECC. Bose–Chaudhuri–Hocquenghem (BCH) codes allow for the correction of multiple bit errors [6, 66, 109, 168], and are used to correct the errors observed during a *single* read from the NAND flash memory [109]. Low-density parity-check (LDPC) codes employ information accumulated over *multiple* read operations to determine the likelihood of each cell containing a bit value 1 or a bit value 0 [55, 119, 168], providing stronger protection at the cost of greater decoding latency and storage overhead [184, 207].

In this section, we briefly overview how an SSD performs error correction when reading data. We first go through an example error correction flow for an SSD that uses either BCH codes (Section 6.1) or LDPC codes (Section 6.2). Next, we compare the error correction strength (i.e., the number of errors that ECC can correct) when we employ BCH codes or LDPC codes in an SSD (Section 6.3). Then, we discuss techniques that can rescue data from an SSD when the BCH/LDPC decoding fails to correct all errors (Section 6.4).

### 6.1 Error Correction Flow With BCH Codes

The SSD starts a read operation by using the initial read reference voltages ($V_{initial}$; see Section 5.5) to read the raw data stored within a page of NAND flash memory into the controller. Once the raw data is read, the controller starts error correction. We first look at the error correction flow using BCH codes [6, 66, 109, 168]. An example flow of the stages for BCH decoding is listed in Algorithm 1, and is shown on the left-hand side of Figure 30a. In the first stage, the ECC engine performs BCH decoding on the raw data, which reports the total number of bit errors in the data. If the data cannot be corrected by the implemented BCH codes, many controllers invoke read-retry (Section 5.4) or read reference voltage optimization (Section 5.5) to find a new set of read reference voltages ($V_{ref}$) that lower the raw bit error rate of the data from the error rate when using $V_{initial}$. The controller uses the new $V_{ref}$ values to read the data again, and then repeats the BCH decoding. BCH decoding is *hard decoding*, where the ECC engine can only use the *hard* bit value information (i.e., either a 1 or a 0) read for a cell using a *single* set of read reference voltages.

If the controller exhausts the maximum number of read attempts (specified as a parameter in the controller), it employs correction techniques such as neighbor-cell-assisted correction (NAC; see Section 5.2) to further reduce the error rate, as shown in the second BCH stage of Algorithm 1. If NAC cannot successfully read the data, the controller then tries to correct the errors using the more expensive superpage-level parity recovery (Section 2.3). The steps for superpage-level parity recovery are shown in the third stage of Algorithm 1. If the data can be extracted successfully from the other pages in the superpage, the data from the target page can be recovered. Whenever data is successfully decoded or recovered, the data is sent to the host (and it is also reprogrammed into a new physical page to ensure that the *corrected* data values are stored for the logical page). Otherwise, the SSD controller reports an uncorrectable error to the host.

### 6.2 Error Correction Flow With LDPC Codes

Figure 30 compares the error correction flow with BCH codes (discussed in Section 6.1) to the flow with LDPC codes. LDPC decoding

**Algorithm 1** Example BCH/LDPC Error Correction Procedure

**First Stage: BCH/LDPC Hard Decoding**

```
Controller gets stored V_initial values to use as V_ref
Flash chips read page using V_ref
ECC decoder decodes BCH/LDPC
if ECC succeeds then
  Controller sends data to host; exit algorithm
else if number of stage iterations not exceeded then
  Controller invokes V_ref optimization to find new V_ref;
           repeats first stage
end
```

**Second Stage (BCH only): NAC**

```
Controller reads immediately-adjacent wordline W
while ECC fails and all possible voltage states for
              adjacent wordline not yet tried do
  Controller goes to next neighbor voltage state V
  Controller sets V_ref based on neighbor voltage state V
  Flash chips read page using V_ref
  Controller corrects cells adjacent to W's cells that
           were programmed to V
  ECC decoder decodes BCH
  if ECC succeeds then
    Controller sends data to host; exit algorithm
    end
end
```

**Second Stage (LDPC only): Level X LDPC Soft Decoding**

```
while ECC fails and X < maximum level N do
  Controller selects optimal value of V_ref^X
  Flash chips do read-retry using V_ref^X
  Controller recomputes LLR_X^{R0} to LLR_X^{RX}
  ECC decoder decodes LDPC
  if ECC succeeds then
    Controller sends data to host; exit algorithm
  else
    Controller goes to soft decoding level X + 1
    end
end
```

**Third Stage: Superpage–Level Parity Recovery**

```
Flash chips read all other pages in the superpage
Controller XORs all other pages in the superpage
if data extraction succeeds then
  Controller sends data to host
else
  Controller reports uncorrectable error
end
```

consists of three major steps. First, the SSD controller performs LDPC hard decoding, where the controller reads the data using the optimal read reference voltages. The process for LDPC hard decoding is similar to that of BCH hard decoding (as shown in the first stage of Algorithm 1), but does not typically invoke read-retry if the first read attempt fails. Second, if LDPC hard decoding cannot correct all of the errors, the controller uses LDPC *soft decoding* to decode the data (which we describe in detail below). Third, if LDPC soft decoding also cannot correct all of the errors, the controller invokes superpage-level parity.

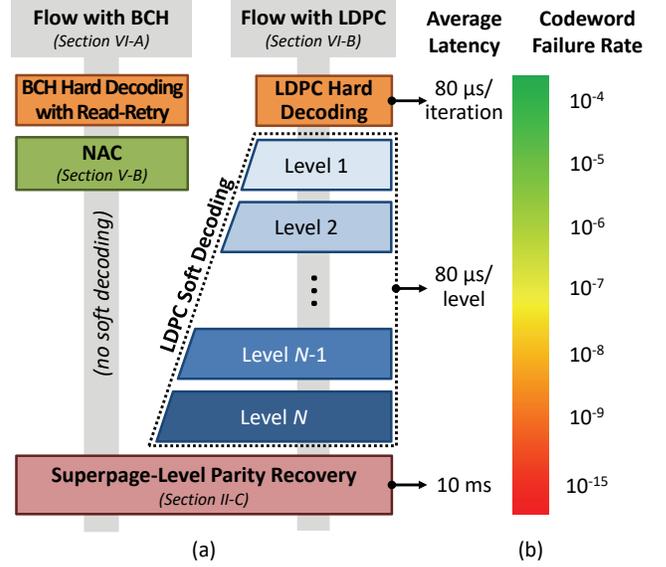

Figure 30: (a) Example error correction flow using BCH codes and LDPC codes. (b) The corresponding average latency and codeword failure rate for each LDPC stage.

**Soft Decoding.** Unlike BCH codes, which require the invocation of expensive superpage-level parity recovery immediately if the hard decoding attempts (BCH hard decoding with read-retry or NAC) fail to return correct data, LDPC decoding fails more gracefully: it can perform multiple levels of *soft decoding* (the second stage in Algorithm 1) after hard decoding fails before invoking superpage-level parity recovery [184, 207]. The key idea of soft decoding is use *soft* information for each cell (i.e., the *probability* that the cell contains a 1 or a 0) obtained from *multiple* reads of the cell via the use of different sets of read reference voltages [55, 119, 168]. Soft information is typically represented by the *log likelihood ratio* (LLR), i.e., the probability of a certain bit being 0, i.e., $P(x = 0|V_{th})$, over the probability of the bit being 1, i.e., $P(x = 1|V_{th})$, given a certain threshold voltage range ($V_{th}$) bounded by two threshold voltage values (i.e., the maximum and the minimum voltage of the threshold voltage range) [184, 207]:

$$\text{LLR} = \log \frac{P(x = 0|V_{th})}{P(x = 1|V_{th})} \quad (10)$$

Every additional level of soft decoding (i.e., the use of a new set of read reference voltages, which we call $V_{ref}^X$ for level $X$) increases the strength of the error correction, as the level *adds* new information about the cell (as opposed to hard decoding, where a new decoding step simply *replaces* prior information about the cell). The new read reference voltages, unlike the ones used for hard decoding, are optimized such that the amount of useful information (or *mutual information*) provided to the LDPC decoder is maximized [184].

Thus, the use of soft decoding reduces the frequency at which superpage-level parity needs to be invoked.

Figure 31 illustrates the read reference voltages used during the first three levels of LDPC soft decoding. At each level, a new read reference voltage is applied, which divides an existing threshold voltage range into two ranges. Based on the bit values read using the various read reference voltages, the SSD controller bins each cell into a certain $V_{th}$ range, and sends the bin categorization of all the cells to the LDPC decoder. For each cell, the decoder applies an LLR value, precomputed by the SSD manufacturer, which corresponds to the cell's bin and decodes the data. For example, as shown in the bottom of Figure 31, the three read reference voltages in Level 3 soft decoding form four threshold voltage ranges (i.e., R0–R3). Each of these ranges corresponds to a different LLR value (i.e., $\text{LLR}_3^{R0}$ to $\text{LLR}_3^{R3}$, where $\text{LLR}_i^{Rj}$ is the LLR value for range $R_j$ in level $i$). Compared with Level 1 soft decoding (shown at the top of Figure 31), which only has two LLR values, Level 3 soft decoding provides more accurate information to the decoder, and thus has stronger error correction capability.

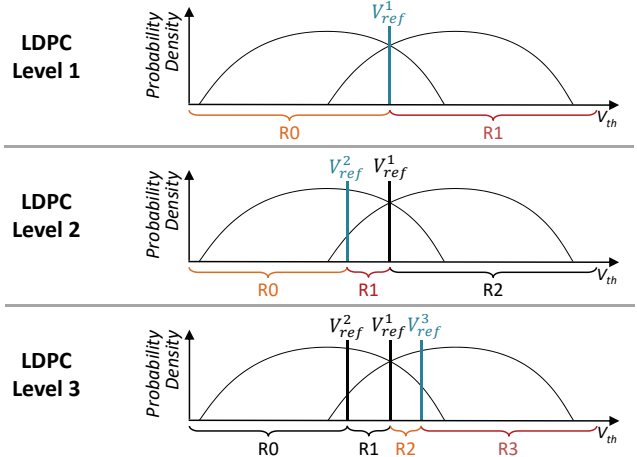

Figure 31: First three levels of LDPC soft decoding, showing the $V_{ref}$ value added at each level, and the resulting threshold voltage ranges (R0–R3) used for flash cell categorization.

**Computing LLR Values.** There are several alternatives for how to compute the LLR values. A common approach for LLR computation is to treat a flash cell as a communication channel, where the channel takes an input program signal (i.e., the target threshold voltage for the cell) and outputs an observed signal (i.e., the current threshold voltage of the cell) [15]. The observed signal differs from the input signal due to the various types of NAND flash memory errors. The communication channel model allows us to break down the threshold voltage of a cell into two components: (1) the expected signal; and (2) the additive signal noise due to errors. By enabling the modeling of these two components separately, the communication channel model allows us to estimate the current threshold voltage distribution of each state [15]. The threshold voltage distributions can be used to predict how likely a cell within a certain voltage region is to belong to a particular voltage state.

One popular variant of the communication channel model assumes that the threshold voltage distribution of each state can be modeled as a Gaussian distribution [15]. If we use the mean observed threshold voltage of each state (denoted as $\mu$) to represent the signal, we find that the P/E cycling noise (i.e., the shift in the distribution of threshold voltages due to the accumulation of charge from repeated programming operations; see Section 4.1) can be modeled as *additive white Gaussian noise* (AWGN) [15], which is represented by the standard deviation of the distribution (denoted as $\sigma$). The closed-form AWGN-based model can be used to determine the LLR value for a cell with threshold voltage $y$, as follows:

$$\text{LLR}(y) = \frac{\mu_1^2 - \mu_0^2}{2\sigma^2} + \frac{y(\mu_0 - \mu_1)}{\sigma^2} \quad (11)$$

where $\mu_0$ and $\mu_1$ are the mean threshold voltages for the distributions of the threshold voltage states for bit value 0 and bit value 1, respectively, and $\sigma$ is the standard deviation of both distributions (assuming that the standard deviation of each threshold voltage state distribution is equal). Since LDPC soft decoding uses threshold voltage ranges to categorize a flash cell, we can substitute $\mu_{R_j}$, the mean threshold voltage of the threshold voltage range $R_j$, in place of $y$ in Equation 11.

The AWGN-based LLR model in Equation 11 provides only an estimate of the LLR, because (1) the actual threshold voltage distributions observed in NAND flash memory are *not* perfectly Gaussian in nature [15, 116]; (2) the controller uses the mean voltage of the threshold voltage range to *approximate* the actual threshold voltage of a cell; and (3) the standard deviations of each threshold voltage state distribution are *not* perfectly equal (see Tables 4–6 in the Appendix). A number of methods have been proposed to improve upon the AWGN-based LLR estimate by: (1) using nonlinear transformations to convert the AWGN-based LLR into a more accurate LLR value [195]; (2) scaling and rounding the AWGN-based LLR to compensate for the estimation error [194]; (3) initially using the AWGN-based LLR to read the data, and, if the read fails, using the ECC information from the failed read attempt to optimize the LLR and to perform the read again with the optimized LLR [43]; and (4) using online and offline training to empirically determine the LLR values under a wide range of conditions (e.g., P/E cycle count, retention time, read disturb count) [196]. The SSD controller can either compute the LLR values at runtime, or statically store precomputed LLR values in a table.

**Determining the Number of Soft Decoding Levels.** If the final level of soft decoding, i.e., level $N$ in Figure 30a, fails, the controller attempts to read the data using superpage-level parity (Section 2.3). The number of levels used for soft decoding depends on the improved reliability that each additional level provides, taking into account the latency of performing additional decoding. Figure 30b shows a rough estimation of the average latency and the codeword failure rate for each stage. There is a tradeoff between the number of levels employed for soft decoding and the expected read latency. For a smaller number of levels, the additional reliability can be worth the latency penalty. For example, while a five-level soft decoding step requires up to 480 μs, it effectively reduces the codeword failure rate by five orders of magnitude. This not only improves overall reliability, but also reduces the frequency of triggering expensive superpage-level parity recovery, which can take around 10 ms [62]. However, manufacturers limit the number of levels, as the benefit of employing an additional soft decoding level (which requires more read operations) becomes smaller due to diminishing returns in the number of additional errors corrected.

## 6.3 BCH and LDPC Error Correction Strength

BCH and LDPC codes provide different strengths of error correction. While LDPC codes can offer a stronger error correction capability, soft LDPC decoding can lead to a greater latency for error correction. Figure 32 compares the error correction strength of BCH codes, hard LDPC codes, and soft LDPC codes [61]. The x-axis shows the raw bit error rate (RBER) of the data being corrected, and the y-axis shows the *uncorrectable bit error rate* (UBER), or the error rate after correction, once the error correction code has been applied. The UBER is defined as the ECC codeword (see Section 2.3) failure rate divided by the codeword length [72]. To ensure a fair comparison, we choose a similar codeword length for both BCH and LDPC codes, and use a similar coding rate (0.935 for BCH, and 0.936 for LDPC) [61]. We make two observations from Figure 32.

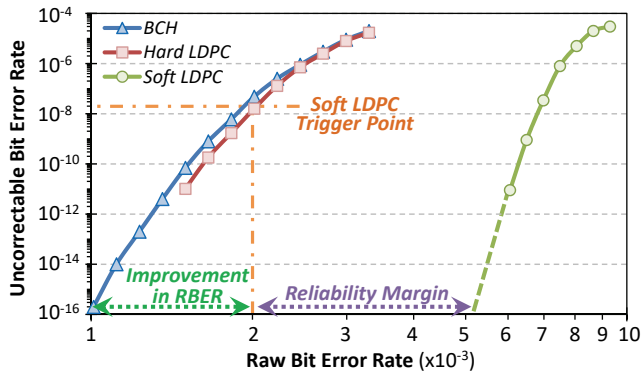

Figure 32: Raw bit error rate versus uncorrectable bit error rate for BCH codes, hard LDPC codes, and soft LDPC codes.

First, we observe that the error correction strength of the hard LDPC code is similar to that of the BCH codes. Thus, on its own, hard LDPC does not provide a significant advantage over BCH codes, as it provides an equivalent degree of error correction with similar latency (i.e., one read operation). Second, we observe that soft LDPC decoding provides a significant advantage in error correction capability. Contemporary SSD manufacturers target a UBER of $10^{-16}$ [72]. The example BCH code with a coding rate of 0.935 can successfully correct data with an RBER of $1.0 \times 10^{-3}$ while remaining within the target UBER. The example LDPC code with a coding rate of 0.936 is more successful with soft decoding, and can correct data with an RBER as high as $5.0 \times 10^{-3}$ while remaining within the target UBER, based on the error rate extrapolation shown in Figure 32. While soft LDPC can tolerate up to five times the raw bit errors as BCH, this comes at a cost of latency (not shown on the graph), as soft LDPC can require several additional read operations after hard LDPC decoding fails, while BCH requires only the original read.

To understand the benefit of LDPC codes over BCH codes, we need to consider the combined effect of hard LDPC decoding and soft LDPC decoding. As discussed in Section 6.2, soft LDPC decoding is invoked *only when hard LDPC decoding fails*. To balance error correction strength with read performance, SSD manufacturers can require that the hard LDPC failure rate cannot exceed a certain threshold, and that the overall read latency (which includes the error correction time) cannot exceed a certain target [61, 62]. For example, to limit the impact of error correction on read performance, a manufacturer can require 99.99% of the error correction operations to be completed after a single read. To meet our example requirement, the hard LDPC failure rate should not be greater than $10^{-4}$ (i.e., 99.99%), which corresponds to an RBER of $2.0 \times 10^{-3}$ and a UBER of $10^{-8}$ (shown as *Soft LDPC Trigger Point* in Figure 32). For only the data that contains one or more failed codewords, soft LDPC is invoked (i.e., soft LDPC is invoked only 0.01% of the time). For our example LDPC code with a coding rate of 0.936, soft LDPC decoding is able to correct these codewords: for an RBER of $2.0 \times 10^{-3}$, using soft LDPC results in a UBER well below $10^{-16}$, as shown in Figure 32.

To gauge the combined effectiveness of hard and soft LDPC codes, we calculate the overhead of using the combined LDPC decoding over using BCH decoding. If 0.01% of the codeword corrections fail, we can assume that in the worst case, each failed codeword resides in a different flash page. As the failure of a single codeword in a flash page causes soft LDPC to be invoked for the entire flash page, our assumption maximizes the number of flash pages that require soft LDPC decoding. For an SSD with four codewords per flash page, our assumption results in up to 0.04% of the data reads requiring soft LDPC decoding. Assuming that the example soft LDPC decoding requires seven additional reads, this corresponds to 0.28% more reads when using combined hard and soft LDPC over BCH codes. Thus, with a 0.28% overhead in the number of reads performed, the combined hard and soft LDPC decoding provides twice the error correction strength of BCH codes (shown as *Improvement in RBER* in Figure 32).

In our example, the lifetime of an SSD is limited by both the UBER and whether more than 0.01% of the codeword corrections invoke soft LDPC, to ensure that the overhead of error correction does not significantly increase the read latency [61]. In this case, when the lifetime of the SSD ends, we can still read out the data correctly from the SSD, albeit at an increased read latency. This is because even though we capped the SSD lifetime to an RBER of $2.0 \times 10^{-3}$ in our example shown in Figure 32, soft LDPC is able to correct data with an RBER as high as $5.0 \times 10^{-3}$ while still maintaining an acceptable UBER ($10^{-16}$) based on the error rate extrapolation shown. Thus, LDPC codes have a margin, which we call the *reliability margin* and show in Figure 32. This reliability margin enables us to trade off lifetime with read latency.

We conclude that with a combination of hard and soft LDPC decoding, an SSD can offer a significant improvement in error correction strength over using BCH codes.

## 6.4 SSD Data Recovery

When the number of errors in data exceeds the ECC correction capability and the error correction techniques in Sections 6.1 and 6.2 are unable to correct the read data, then data loss can occur. At this point, the SSD is considered to have reached the end of its lifetime. In order to avoid such data loss and *recover* (or, *rescue*) the data from the SSD, we can harness our understanding of data retention and read disturb behavior. The SSD controller can employ two conceptually similar mechanisms, *Retention Failure Recovery* (RFR) [17] and *Read Disturb Recovery* (RDR) [16], to undo errors that were introduced into the data as a result of data retention and read disturb, respectively. The key idea of both of these mechanisms is to exploit the wide variation of different flash cells in their susceptibility to data retention loss and read disturbance effects, respectively, in order to correct some of the errors *without* the assistance of ECC so that the remaining error count falls within the ECC error correction capability.

When a flash page read fails (i.e., uncorrectable errors exist), RFR and RDR record the current threshold voltages of each cell in the page using the read-retry mechanism (see Section 5.4), and identify the cells that are *susceptible* to generating errors due to retention and read disturb (i.e., cells that lie at the tails of the threshold voltage distributions of each state, where the distributions overlap with each other), respectively. We observe that some flash cells are more likely to be affected by retention leakage and read disturb than others, as a result of process variation [16, 17]. We call these cells retention/read disturb *prone*, while cells that are less likely to be affected are called retention/read disturb *resistant*. RFR and RDR classify the susceptible cells as retention/read disturb prone or resistant by inducing *even more* retention and read disturb on the failed flash page, and then recording the new threshold voltages of the susceptible cells. We classify the susceptible cells by observing the magnitude of the threshold voltage shift due to the additional retention/read disturb induction.

Figure 33 shows how the threshold voltage of a retention-prone cell (i.e., a *fast-leaking* cell, labeled P in the figure) decreases over time (i.e., the cell shifts to the left) due to retention leakage, while the threshold voltage of a retention- resistant cell (i.e., a *slow-leaking* cell, labeled R in the figure) does not change significantly over time. Retention Failure Recovery (RFR) uses this classification of retention- prone versus retention-resistant cells to correct the data from the failed page *without* the assistance of ECC. Without loss of generality, let us assume that we are studying susceptible cells near the intersection of two threshold voltage distributions X and Y, where Y contains higher voltages than X. Figure 33 highlights the region of cells considered susceptible by RFR using a box, labeled *Susceptible*. A susceptible cell within the box that is retention prone likely belongs to distribution Y, as a retention-prone cell shifts rapidly to a lower voltage (see the circled cell labeled P within the *susceptible* region in the figure). A retention-resistant cell in the same *susceptible* region likely belongs to distribution X (see the boxed cell labeled R within the *susceptible* region in the figure).

Similarly, Read Disturb Recovery (RDR) uses the classification of read disturb prone versus read disturb resistant cells to correct data. For RDR, disturb-prone cells shift more rapidly to higher voltages, and are thus likely to belong to distribution X, while disturb-resistant cells shift little and are thus likely to belong to distribution Y. Both RFR and RDR correct the bit errors for the susceptible cells based on such *expected* behavior, reducing the number of errors that ECC needs to correct.

RFR and RDR are highly effective at reducing the error rate of failed pages, reducing the raw bit error rate by 50% and 36%, respectively, as shown in our prior works [16, 17], where more detailed information and analyses can be found.

## 7 EMERGING RELIABILITY ISSUES FOR 3D NAND FLASH

Recently, manufacturers have begun to produce SSDs that contain three-dimensional (3D) NAND flash memory, where multiple layers are vertically stacked to increase the density and to improve the scalability of the memory [205]. Instead of using floating gate transistors, which store charge on a conductor, most 3D NAND flash memories currently use *charge trap transistors*, which use insulating material to store charge. While the high-level behavior of charge trap transistors is similar to FG transistors, charge trap transistors do introduce some differences in terms of reliability for 3D NAND flash (as opposed to 2D *planar* NAND flash, which we have examined throughout this article so far). For example, the tunneling oxide in charge trap transistors is *less* susceptible to breakdown than the oxide in floating gate transistors during high-voltage operation, increasing the endurance of the transistor [205]. Charge trap transistors are, however, *more* susceptible to data retention leakage. Due to the possibility that charge can now escape (i.e., migrate) across the z-dimension in addition to through the tunnel oxide, 3D NAND flash cells tend to leak more rapidly, especially soon after being programmed [205].

Another, albeit short-term, change with 3D NAND flash is the *increase* in process technology feature size. Contemporary 3D NAND flash can contain 48–64 layers, allowing manufacturers to use larger feature sizes (e.g., 50–54 nm) than commonly used feature sizes in planar flash (e.g., 15–19 nm while still increasing memory density [205]. As discussed in Section 4, many of the errors observed in 2D planar NAND flash are exacerbated as a result of significant process scaling. For example, while read disturb is a prominent source of errors at small feature sizes (e.g., 20–24 nm), its effects are small at larger feature sizes [16]. Likewise, cell-to-cell program interference is not a significant issue at larger process technologies, leading manufacturers to revert to one-shot programming (see Section 3.4) for 3D NAND flash [152]. As the transistors are larger in the current 3D NAND flash generations, the endurance (i.e., the maximum P/E cycle count) of the flash cells has increased as well, by over an order of magnitude [152]. However, rigorous studies that examine error characteristics of and error mitigation techniques for 3D NAND flash memories are yet to be published.

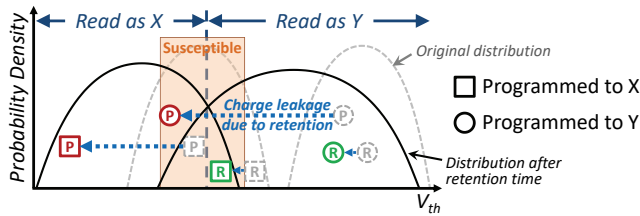

Figure 33: Some retention-prone (P) and retention-resistant (R) cells are incorrectly read after charge leakage due to retention time. RFR identifies and corrects the incorrectly read cells based on their leakage behavior.

While these changes with 3D NAND flash are likely to reduce reliability issues due to program interference and read disturb as compared to planar NAND flash, the other errors outlined in Section 4 are likely to remain prevalent in 3D NAND flash. In fact, retention errors are likely to become exacerbated. As such, all described techniques covered in this paper still apply to 3D NAND flash, though their relative benefits are yet to be evaluated. With its increased susceptibility to data retention leakage, advanced retention mitigation and recovery techniques, such as those described in Sections 5.3 and 5.5, should be even more actively developed and investigated for 3D NAND flash memory. Furthermore, 3D NAND flash memory is expected to scale down to smaller process technologies in the coming years, reaching the feature sizes of modern planar flash memory, and to make use of FG transistors [205], just like modern planar flash memory. As such, with technology scaling of 3D NAND flash memory, we can expect that all of the reliability issues highlighted in this paper will be exhibited in SSDs that utilize 3D NAND flash memory.

## 8 SIMILAR ERRORS IN OTHER MEMORY TECHNOLOGIES

As we discussed in Section 4, there are five major sources of errors in flash-memory-based SSDs. Many of these error sources can also be found in other types of memory and storage technologies. In this section, we take a brief look at the major reliability issues that exist within DRAM and in emerging nonvolatile memories. In particular, we focus on DRAM in our discussion, as modern SSD controllers have access to dedicated DRAM of considerable capacity (e.g., 1 GB for every 1 TB of SSD capacity), which exists within the SSD package (see Section 2). Major sources of errors in DRAM include data retention, cell-to-cell interference, and read disturb. There is a wide body of work on mitigation mechanisms for the errors we describe in this section, but we explicitly discuss only a select number of them here.

**Data Retention Errors in DRAM.** DRAM uses the charge within a capacitor to represent one bit of data. Much like the floating gate within NAND flash memory, charge leaks from the DRAM capacitor over time, leading to data retention issues. Charge leakage in DRAM, if left unmitigated, can lead to much more rapid data loss than the leakage observed in a NAND flash cell. While leakage from a NAND flash cell typically leads to data loss after several days to years of retention time (see Section 4.4), leakage from a DRAM cell leads to data loss after a retention time on the order of *milliseconds* to *seconds* [112]. Due to the rapid charge leakage from DRAM cells, a DRAM controller periodically refreshes all DRAM cells in place [30, 73, 80, 112, 113, 154, 158] (similar to the techniques discussed in Section 5.3, but at a much smaller time scale). DRAM standards require a DRAM cell to be refreshed once every 64 ms [73]. As the density of DRAM continues to increase over successive product generations (e.g., by 128x between 1999 and 2017 [29, 32]), the performance and energy overheads required to refresh an entire DRAM module have grown significantly [113].

To combat the growing performance and energy overheads of refresh, two classes of techniques have been developed. The first class of techniques reduce the *frequency* of refresh operations without sacrificing the reliability of data stored in DRAM (e.g., [71, 80, 82, 113, 154, 158, 182]). To reduce the frequency of refresh operations, a number of works take advantage of the fact that the vast majority of DRAM cells can retain data without loss for much longer than 64 ms, as various experimental studies of real DRAM chips (e.g., [80, 87, 106, 113, 154, 158]) demonstrate. The second class of techniques reduce the interference caused by refresh requests on demand requests (e.g., [30, 133, 172]). These works either change the scheduling order of refresh requests [30, 133, 172] or slightly modify the DRAM architecture to enable the servicing of refresh and demand requests in parallel [30]. More findings on the nature of DRAM data retention and associated errors, as well as relevant experimental data from modern DRAM chips, can be found in our prior works [29, 30, 63, 80, 81, 83, 106, 112, 113, 154, 158].

**Cell-to-Cell Interference Errors in DRAM.** Another similarity between the capacitive DRAM cell and the floating gate cell in NAND flash memory is that they are both vulnerable to cell-to-cell interference. In DRAM, one important way in which cell-to-cell interference exhibits itself is the data-dependent retention behavior, where the retention time of a DRAM cell is dependent on the values written to *nearby* DRAM cells [80–82, 112, 154]. This phenomenon is called *data pattern dependence* (DPD) [112]. Data pattern dependence in DRAM is similar to the data-dependent nature of program interference that exists in NAND flash memory (see Section 4.3). Within DRAM, data dependence occurs as a result of parasitic capacitance coupling (between DRAM cells). Due to this coupling, the amount of charge stored in one cell's capacitor can inadvertently affect the amount of charge stored in an adjacent cell's capacitor [80–82, 112, 154]. As DRAM cells become smaller with technology scaling, cell-to-cell interference worsens because parasitic capacitance coupling between cells increases [80, 112]. More findings on cell-to-cell interference and the data-dependent nature of cell retention times in DRAM, along with experimental data obtained from modern DRAM chips, can be found in our prior works [29, 80–83, 112, 154, 158].

**Read Disturb Errors in DRAM.** Commodity DRAM chips that are sold and used in the field today exhibit read disturb errors [94], also called *RowHammer*-induced errors [136], which are *conceptually* similar to the read disturb errors found in NAND flash memory (see Section 4.5). Repeatedly accessing the same row in DRAM can cause bit flips in data stored in adjacent DRAM rows. In order to access data within DRAM, the row of cells corresponding to the requested address must be *activated* (i.e., opened for read and write operations). This row must be *precharged* (i.e., closed) when another row in the same DRAM bank needs to be activated. Through experimental studies on a large number of real DRAM chips, we show that when a DRAM row is activated and precharged repeatedly (i.e., *hammered*) enough times within a DRAM refresh interval, one or more bits in physically-adjacent DRAM rows can be flipped to the wrong value [94]. This DRAM failure mode affects more than 80% of the DRAM chips we tested [94]. As indicated above, this read disturb error mechanism in DRAM is popularly called RowHammer [136].

Various recent works show that RowHammer can be maliciously exploited by user-level software programs to (1) induce errors in existing DRAM modules [94, 136] and (2) launch attacks to compromise the security of various systems [7, 9, 57, 136, 160, 164, 165, 181, 198]. For example, by exploiting the RowHammer read disturb mechanism, a user-level program can gain kernel-level privileges on real laptop systems [164, 165], take over a server vulnerable to RowHammer [57], take over a victim virtual machine running on the same system [7], and take over a mobile device [181]. Thus, the RowHammer read disturb mechanism is a prime (and perhaps the first) example of how a circuit-level failure mechanism in DRAM can cause a practical and widespread system security vulnerability.[3] We believe similar (yet more difficult to exploit) vulnerabilities exist in MLC NAND flash memory as well, as described in our recent work [12].

The RowHammer effect in DRAM worsens as the manufacturing process scales down to smaller node sizes [94, 136]. More findings on RowHammer, along with extensive experimental data from real DRAM devices, can be found in our prior works [89, 94, 136].

**Large-Scale DRAM Error Studies.** Like flash memory, DRAM is employed in a wide range of computing systems, at scale. Thus, there is a similar need to study the aggregate behavior of errors observed in a large number of DRAM chips deployed in the field. Akin to the large-scale flash memory SSD reliability studies discussed in Section 4.6, a number of experimental studies characterize the reliability of DRAM at large scale in the field (e.g., [70, 123, 163, 169, 170]). Two notable results from these studies are that (1) unlike SSDs, DRAM does not show any clearly discernable trend where higher utilization and age lead to a greater raw bit error rate [123]; and (2) the increase in the density of DRAM chips with technology scaling leads to higher error rates [123].

**Latency-Related Errors in DRAM.** Other experimental studies examine the tradeoff between DRAM reliability and latency [28, 29, 32, 33, 63, 101, 103, 106]. These works perform extensive experimental studies on real DRAM chips to identify the effect of (1) temperature, (2) supply voltage, and (3) manufacturing process variation that exists in DRAM on the latency and reliability characteristics of different DRAM cells and chips. The temperature, supply voltage, and manufacturing process variation all dictate the amount of time that each cell needs to safely complete its operations. Our works examine how one can reliably exploit (1) latency variation across different operating temperatures and across different DRAM modules to reduce the access latency of each module [106]; (2) the relation between supply voltage and latency variation to reduce the amount of system energy consumed [33]; and (3) manufacturing process induced latency variation [32] and design-induced latency variation [103] across the cells within a single DRAM chip to reduce access latency to different parts of the chip. One can further reduce latency by sacrificing some amount of reliability and performing error correction to fix the resulting errors [103]. More information about the errors caused by reduced latency operation in DRAM chips and the tradeoff between reliability and latency can be found in our prior works [29, 32, 33, 63, 101, 103, 106, 118].

**Error Correction in DRAM.** In order to protect the data stored within DRAM from various types of errors, some (but not all) DRAM modules employ ECC [118]. The ECC employed within DRAM is much weaker than the ECC employed in SSDs (see Section 6) for various reasons. First, DRAM has a much lower access latency, and error correction mechanisms should be designed to ensure that DRAM access latency does not increase significantly. Second, the error rate of a DRAM chip tends to be lower than that of a flash memory chip. Third, the granularity of access is much smaller in a DRAM chip than in a flash memory chip, and hence sophisticated error correction can come at a high cost. The most common ECC algorithm used in commodity DRAM modules is *SECDED* (single error correction, double error detection) [118]. Another ECC algorithm available for some commodity DRAM modules is *Chipkill*, which can tolerate the failure of an entire DRAM chip within a module [47]. For both SECDED and Chipkill, the ECC information is stored on one or more extra chips within the DRAM module, and, on a read request, this information is sent alongside the data to the memory controller, which performs the error detection and correction algorithm.

As DRAM scales to smaller technology nodes, its error rate continues to increase [94, 123, 135, 136, 139] and effects like read disturb [94], cell-to-cell interference [80–83, 112, 154], and variable retention time [80, 112, 154, 158] become more severe [94, 135, 136, 139]. As a result, there is an increasing need for (1) employing ECC algorithms in *all* DRAM chips/modules; (2) developing more sophisticated and efficient ECC algorithms for DRAM chips/modules; and (3) developing error-specific mechanisms for error correction. To this end, recent work follows various directions. First, in-DRAM ECC, where correction is performed within the DRAM module itself (as opposed to in the controller), is proposed [79]. One work shows how exposing this in-DRAM ECC information to the memory controller can provide Chipkill-like error protection at much lower overhead than the traditional Chipkill mechanism [141]. Second, various works explore and develop stronger ECC algorithms for DRAM (e.g., [85, 86, 188]), and explore how to make ECC more efficient based on the current DRAM error rate (e.g., [2, 38, 47, 180]). Third, recent work shows how the cost of ECC protection can be reduced by (1) exploiting *heterogeneous reliability memory* [118], where different portions of DRAM use different strengths of error protection based on the error tolerance of different applications and different types of data [114, 118], and (2) using the additional DRAM capacity that is otherwise used for ECC to improve system performance when reliability is not as important for the given application and/or data [117].

Many of these works that propose error mitigation mechanisms do not distinguish between the characteristics of different types of errors. We believe that in addition to providing sophisticated and efficient ECC mechanisms in DRAM, there is also significant value in and opportunity for exploring *specialized* error mitigation mechanisms that are *customized for different error types*, again, just as it is done for flash memory (as we discussed in Section 5). One such example of a specialized error mitigation mechanism is targeted to fix the RowHammer read disturb mechanism, and is called *Probabilistic Adjacent Row Activation* (PARA) [94, 136]. The key idea of PARA is to refresh the rows that are physically adjacent to an activated row, with a very low probability. PARA is shown to be very effective in fixing the RowHammer problem at no storage cost and at very low performance overhead [94].

---

[3]Note that various solutions to RowHammer exist [89, 94, 136], but we do not discuss them here.

**Errors in Emerging Nonvolatile Memory Technologies.**
DRAM operations are several orders of magnitude faster than SSD operations, but DRAM has two major disadvantages. First, DRAM offers orders of magnitude less storage density than NAND-flash-memory-based SSDs. Second, DRAM is volatile (i.e., the stored data is lost on a power outage). Emerging nonvolatile memories, such as *phase-change memory* (PCM) [97–99, 159, 192, 204, 208], *spin-transfer torque magnetic RAM* (STT-RAM or STT-MRAM) [96, 140], *metal-oxide resistive RAM* (RRAM) [191], and *memristors* [39, 171], are expected to bridge the gap between DRAM and SSDs, providing DRAM-like access latency and energy, and at the same time SSD-like large capacity and nonvolatility (and hence SSD-like data persistence). PCM-based devices are expected to have a limited lifetime, as PCM can only endure a certain number of writes [97, 159, 192], similar to the P/E cycling errors in NAND-flash-memory-based SSDs (though PCM's write endurance is higher than that of SSDs). PCM suffers from resistance drift [192], where the resistance used to represent the value shifts higher over time (and eventually introduces a bit error), similar to how charge leakage in NAND flash memory and DRAM lead to retention errors over time. STT-RAM predominantly suffers from retention failures, where the magnetic value stored for a single bit can flip over time, and read disturb (different from the read disturb in DRAM and flash memory), where reading a bit in STT-RAM can inadvertently induce a write to that same bit [140]. Due to the nascent nature of emerging nonvolatile memory technologies and the lack of availability of large-capacity devices built with them, extensive and dependable experimental studies have yet to be conducted on the reliability of real PCM, STT-RAM, RRAM, and memristor chips. However, we believe that similar error mechanisms to those we discussed in this paper for flash memory and DRAM are likely to be prevalent in emerging technologies as well, albeit with different underlying mechanisms and error rates.

## 9 CONCLUSION

We provide a survey of the fundamentals of and recent research in NAND-flash-memory-based SSD reliability. As the underlying NAND flash memory within SSDs scales to increase storage density, we find that the rate at which raw bit errors occur in the memory increases significantly, which in turn reduces the lifetime of the SSD. We describe the prevalent error mechanisms that affect NAND flash memory, and examine how they behave in modern NAND flash memory chips. To compensate for the increased raw bit error rate with technology scaling, a wide range of error mitigation and data recovery mechanisms have been proposed. These techniques effectively undo some of the SSD lifetime reductions that occur due to flash memory scaling. We describe the state-of-the-art techniques for error mitigation and data recovery, and discuss their benefits. Even though our focus is on MLC and TLC NAND flash memories, for which we provide data from real flash chips, we believe that these techniques will be applicable to emerging 3D NAND flash memory technology as well, especially when the process technology scales to smaller nodes. Thus, we hope the tutorial presented in this work on fundamentals and recent research not only enables practitioners to get acquainted with flash memory errors and how they are mitigated, but also helps inform future directions in NAND flash memory and SSD development as well as system design using flash memory. We believe future is bright for system-level approaches that codesign system and memory [135, 136, 139] to enhance overall scaling of platforms, and we hope that the examples of this approach presented in this tutorial inspire researchers and developers to enhance future computing platforms via such system-memory codesign.

## ACKNOWLEDGMENTS

The authors would like to thank Rino Micheloni for his helpful feedback on earlier drafts of the paper. They would also like to thank their collaborator Seagate for their continued dedicated support. Special thanks also goes to our research group SAFARI's industrial sponsors over the past six years, especially Facebook, Google, Huawei, Intel, Samsung, Seagate, VMware. This work was also partially supported by ETH Zürich, the Intel Science and Technology Center for Cloud Computing, the Data Storage Systems Center at Carnegie Mellon University, and NSF grants 1212962 and 1320531.

A version of the paper is published as an invited article in *Proceedings of the IEEE* [11]. This version is almost identical to [11].

## APPENDIX: TLC THRESHOLD VOLTAGE DISTRIBUTION DATA

**Table 4:** Normalized mean (top) and standard deviation (bottom) values for threshold voltage distribution of each voltage state at various P/E cycle counts (Section 4.1).

| P/E Cycles | ER | P1 | P2 | P3 | P4 | P5 | P6 | P7 |
|---|---|---|---|---|---|---|---|---|
| 0 | -110.0 | 65.9 | 127.4 | 191.6 | 254.9 | 318.4 | 384.8 | 448.3 |
| 200 | -110.4 | 66.6 | 128.3 | 192.8 | 255.5 | 319.3 | 385.0 | 448.6 |
| 400 | -105.0 | 66.0 | 127.3 | 191.7 | 254.5 | 318.2 | 383.9 | 447.7 |
| 1,000 | -99.9 | 66.5 | 127.1 | 191.7 | 254.8 | 318.1 | 384.4 | 447.8 |
| 2,000 | -92.7 | 66.6 | 128.1 | 191.9 | 254.9 | 318.3 | 384.3 | 448.1 |
| 3,000 | -84.1 | 68.3 | 128.2 | 193.1 | 255.7 | 319.2 | 385.4 | 449.1 |

| P/E Cycles | ER | P1 | P2 | P3 | P4 | P5 | P6 | P7 |
|---|---|---|---|---|---|---|---|---|
| 0 | 45.9 | 9.0 | 9.4 | 8.9 | 8.8 | 8.9 | 9.3 | 8.5 |
| 200 | 46.2 | 9.2 | 9.8 | 9.0 | 8.8 | 9.0 | 9.1 | 8.5 |
| 400 | 46.4 | 9.2 | 9.5 | 9.1 | 8.8 | 8.8 | 9.0 | 8.6 |
| 1,000 | 47.3 | 9.5 | 9.4 | 9.1 | 9.3 | 8.9 | 9.4 | 8.8 |
| 2,000 | 48.2 | 9.7 | 9.7 | 9.4 | 9.3 | 9.1 | 9.5 | 9.1 |
| 3,000 | 49.4 | 10.2 | 10.2 | 9.6 | 9.7 | 9.5 | 9.8 | 9.4 |

**Table 5:** Normalized mean (top) and standard deviation (bottom) values for threshold voltage distribution of each voltage state at various data retention times (Section 4.4).

| Time | ER | P1 | P2 | P3 | P4 | P5 | P6 | P7 |
|---|---|---|---|---|---|---|---|---|
| 1 day | -92.7 | 66.6 | 128.1 | 191.9 | 254.9 | 318.3 | 384.3 | 448.1 |
| 1 week | -86.7 | 67.5 | 128.1 | 191.4 | 253.8 | 316.5 | 381.8 | 444.9 |
| 1 month | -84.4 | 68.6 | 128.7 | 191.6 | 253.5 | 315.8 | 380.9 | 443.6 |
| 3 months | -75.6 | 72.8 | 131.6 | 193.3 | 254.3 | 315.7 | 380.2 | 442.2 |
| 1 year | -69.4 | 76.6 | 134.2 | 195.2 | 255.3 | 316.0 | 379.6 | 440.8 |

| Time | ER | P1 | P2 | P3 | P4 | P5 | P6 | P7 |
|---|---|---|---|---|---|---|---|---|
| 1 day | 48.2 | 9.7 | 9.7 | 9.4 | 9.3 | 9.1 | 9.5 | 9.1 |
| 1 week | 46.4 | 10.7 | 10.8 | 10.5 | 10.6 | 10.3 | 10.6 | 10.6 |
| 1 month | 46.8 | 11.3 | 11.2 | 11.0 | 10.9 | 10.8 | 11.2 | 11.1 |
| 3 months | 45.9 | 12.0 | 11.8 | 11.5 | 11.4 | 11.4 | 11.7 | 11.7 |
| 1 year | 45.9 | 12.8 | 12.4 | 12.0 | 12.0 | 11.9 | 12.3 | 12.4 |

**Table 6:** Normalized mean (top) and standard deviation (bottom) values for threshold voltage distribution of each voltage state at various read disturb counts (Section 4.5).

| Read Disturbs | ER | P1 | P2 | P3 | P4 | P5 | P6 | P7 |
|---|---|---|---|---|---|---|---|---|
| 1 | -84.2 | 66.2 | 126.3 | 191.5 | 253.7 | 316.8 | 384.3 | 448.0 |
| 1,000 | -76.1 | 66.7 | 126.6 | 191.5 | 253.6 | 316.4 | 383.8 | 447.5 |
| 10,000 | -57.0 | 67.9 | 127.0 | 191.5 | 253.3 | 315.7 | 382.9 | 445.7 |
| 50,000 | -33.4 | 69.9 | 128.0 | 191.9 | 253.3 | 315.4 | 382.0 | 444.1 |
| 100,000 | -20.4 | 71.6 | 128.8 | 192.1 | 253.3 | 315.0 | 381.1 | 443.0 |

| Read Disturbs | ER | P1 | P2 | P3 | P4 | P5 | P6 | P7 |
|---|---|---|---|---|---|---|---|---|
| 1 | 48.2 | 9.7 | 9.7 | 9.4 | 9.3 | 9.1 | 9.5 | 9.1 |
| 1,000 | 47.4 | 10.7 | 10.8 | 10.5 | 10.6 | 10.3 | 10.6 | 10.6 |
| 10,000 | 46.3 | 12.0 | 11.7 | 11.4 | 11.4 | 11.4 | 11.7 | 11.7 |
| 50,000 | 46.1 | 12.3 | 12.1 | 11.7 | 11.6 | 11.7 | 12.0 | 12.4 |
| 100,000 | 45.9 | 12.8 | 12.4 | 12.0 | 12.0 | 11.9 | 12.3 | 12.4 |

## ABOUT THE AUTHORS

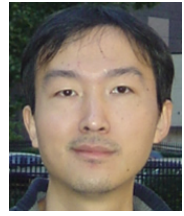

**Yu Cai** received the B.S. degree from Beijing University of Posts and Telecommunications in Telecommunication Engineering, Beijing, China, the M.S. degree in electronic engineering from Tsinghua University, Beijing, China, and the Ph.D. degree in computer engineering from Carnegie Mellon University, Pittsburgh, PA, USA.

He has worked as a solid-state disk system architect at SK Hynix, Seagate Technology, Avago Technologies, and LSI Corporation. Prior to that, he worked on wireless communications at the Hong Kong Applied Science and Technology Research Institute (ASTRI), Alcatel-Lucent, and Microsoft Research Asia (MSRA). He has authored over 20 peer-reviewed papers and holds more than 30 U.S. patents.

Dr. Cai received the Best Paper Runner-Up Award from the IEEE International Symposium on High-Performance Computer Architecture (HPCA) in 2015. He also received the Best Paper Award from the DFRWS Digital Forensics Research Conference Europe in 2017.

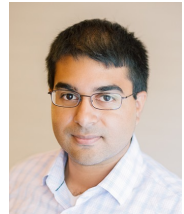

**Saugata Ghose** received dual B.S. degrees in computer science and in computer engineering from Binghamton University, State University of New York, USA, and the M.S. and Ph.D. degrees from Cornell University, Ithaca, NY, USA, where he was the recipient of the NDSEG Fellowship and the ECE Director's Ph.D. Teaching Assistant Award.

He is a Systems Scientist in the Department of Electrical and Computer Engineering at Carnegie Mellon University, Pittsburgh, PA, USA. He is a member of the SAFARI Research Group, led by Dr. Onur Mutlu. His current research interests include application- and system-aware memory and storage systems, flash reliability, architectural solutions for large-scale systems, GPUs, and emerging memory technologies.

Dr. Ghose received the Best Paper Award from the DFRWS Digital Forensics Research Conference Europe in 2017. For more information, see his website at https://ece.cmu.edu/~saugatag/.

**Erich F. Haratsch** is Director of Engineering at Seagate Technology, where he is responsible for the architecture of flash controllers. He leads the development of hardware and firmware features that improve the performance, quality of service, endurance, error correction and media management capabilities of solid-state drives. Earlier in his career, he developed signal processing and error correction technologies for hard disk drive controllers at LSI Corporation and Agere Systems, which shipped in more than one billion chips. He started his engineering career at Bell Labs Research, where he invented new chip architectures for Gigabit Ethernet over copper and optical communications. He is a frequent speaker at leading industry events, is the author of over 40 peer-reviewed journal and conference papers, and holds more than 100 U.S. patents.

He earned his M.S. and Ph.D. degrees in electrical engineering from the Technical University of Munich (Germany).

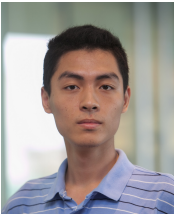

**Yixin Luo** received the B.S.E. degree in computer engineering from the University of Michigan, Ann Arbor, MI, USA and the B.S.E. degree in electrical engineering from Shanghai Jiao Tong University, Shanghai, China, in 2012. He is currently working toward the Ph.D. degree in computer science at Carnegie Mellon University, Pittsburgh, PA, USA.

At Carnegie Mellon, he is involved in research on DRAM and flash reliability, and on datacenter reliability and cost optimization.

Mr. Luo received the Best Paper Award and the Best Paper Runner-Up Award from the IEEE International Symposium on High-Performance Computer Architecture in 2012 and 2015, respectively, and the Best Paper Award from the DFRWS Digital Forensics Research Conference Europe in 2017.

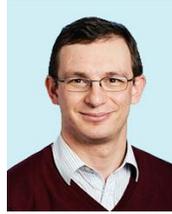

**Onur Mutlu** received B.S. degrees in computer engineering and psychology from the University of Michigan, Ann Arbor, MI, USA and the M.S. and Ph.D. degrees in electrical and computer engineering from the University of Texas at Austin, USA.

He is a Professor of Computer Science at ETH Zürich, Switzerland. He is also a faculty member at Carnegie Mellon University, Pittsburgh, PA, USA, where he previously held the William D. and Nancy W. Strecker Early Career Professorship. His current broader research interests are in computer architecture, systems, and bioinformatics. He is especially interested in interactions across domains and between applications, system software, compilers, and microarchitecture, with a major current focus on memory and storage systems. His industrial experience spans starting the Computer Architecture Group at Microsoft Research (2006–2009), and various product and research positions at Intel Corporation, Advanced Micro Devices, VMware, and Google. His computer architecture course lectures and materials are freely available on YouTube, and his research group makes software artifacts freely available online.

Dr. Mutlu received the inaugural IEEE Computer Society Young Computer Architect Award, the inaugural Intel Early Career Faculty Award, faculty partnership awards from various companies, and a healthy number of best paper and "Top Pick" paper recognitions at various computer systems and architecture venues. For more information, see his webpage at http://people.inf.ethz.ch/omutlu/.